\documentclass[
	aps,
	prl,
	reprint,
	amsmath,amssymb,
	superscriptaddress,floatfix
]{revtex4-1}

\usepackage{graphicx}% Include figure files
\usepackage{dcolumn}% Align table columns on decimal point
\usepackage{bm}% bold math
\usepackage{hyperref}% add hypertext capabilities
\usepackage[mathlines]{lineno}% Enable numbering of text and 

\usepackage{color,soul} 
\usepackage{comment} 	
\usepackage{xcolor} %required for \textcolor command
\usepackage{verbatim} %Reimplementation of and extensions to LaTeX verbatim, also provides a comment environment

\setul{0.3ex}{0.25ex} 
\setulcolor{magenta}

\begin{document}
\title{Quantitative signal extraction in the dynamic range of nanomechanical systems by free and constrained fitting}
\author{Fan Yang}
\affiliation{Department of Physics, Universit{\"a}t Konstanz, 78464 Konstanz, Germany}
\author{Reimar Waitz}
%\altaffiliation[Now at: ]{RATIONAL AG, Iglinger Str. 62, 86899 Landsberg am Lech, Germany}
\author{Mengqi Fu}
 \email{mengqi.fu@uni-konstanz.de}
\affiliation{Department of Physics, Universit{\"a}t Konstanz, 78464 Konstanz, Germany}
\author{Elke Scheer}
\affiliation{Department of Physics, Universit{\"a}t Konstanz, 78464 Konstanz, Germany}

\begin{abstract}
\noindent We present a free and a constrained fitting procedure for quantitative signal extraction of nanomechanical systems in the dynamic range and for physical model testing. We demonstrate that applying the free-fitting procedure to the measured frequency response of silicon nitride (SiN) nanomembranes at varying pressure enables us to disentangle the intrinsic membrane vibration properties from the system response, thereby giving quantitative access to the eigenfrequency, quality factor, coupling strength between resonator and drive system, and to system noise. The validity of physical models for quantities such as excitation, fluctuations, and damping mechanisms can be verified by imposing additional mathematical links between different physical parameters as constraints in the constrained fitting procedure. We verify the performance of the constrained fitting procedure for the same samples tested in various experimental setups.
\end{abstract}


\maketitle

\indent Nanomechanical resonators with high quality factor \cite{beccari2022strained,tsaturyan2017ultracoherent} are extensively used in a variety of applications such as sensors \cite{lemme2020nanoelectromechanical}, quantum technology \cite{barzanjeh2021optomechanics}, and coupling with other physical systems \cite{kovsata2020spin, karg2020light, wang2022enhanced}. As the dimension of the resonators decrease, on the one hand, the vibration amplitude of the resonators becomes smaller leading to a stronger requirement for the detection system \cite{wen2020coherent, zhang2015vibrational} while, on the other hand, the more sensitive nanomechanical system is more prone to perturbations of the environment \cite{steeneken2021dynamics,OpticalcontrolMoS2, gisler2021soft}. Both aspects challenge the improvement of sensitivity of the nanomechanical resonators. To overcome this challenge, one approach is to efficiently reduce the fluctuations caused by the setup as well as the environment. This reduction of course has its limitations and increases the requirements for the instrumentation. Another approach is hence to characterize the system properties including the noise and to disentangle them from the total signal to ensure the accurate acquisition of the intrinsic contribution of the nanomechanical resonator. \\
\indent In this letter, we present a free and a constrained fitting method with customized scripts to decompose the measured frequency response into the contributions of the intrinsic mechanical vibration of resonators and of the system fluctuations.
We demonstrate the procedure for different membrane resonators in varying environment and tested in different measurement schemes, including optical or electromagnetic platforms. Through the presented fitting procedures, the eigenfrequencies, $Q$ factors, force coupling strength and the system noise of the membrane have been quantitatively determined. In addition, the validity of several physical models explaining the dependence of mechanical response of resonators on the gas pressure has been tested by constrained fitting.\\
\indent The membranes are fabricated using wet etching of silicon in aqueous potassium hydroxide (KOH). A 0.5 mm thick commercial (100) silicon wafer which both sides are coated with a ~500 nm thick layer of silicon nitride (SiN) is applied. The backside layer is patterned by laser ablation to form the etch mask. After anisotropic etching process, the parts of the silicon substrate under the openings of the  etch masks are removed. The membranes supported by a massive silicon frame are formed and their sizes can be controlled by varying the size of the openings. The silicon wafer then is cut into pieces (referred to as chips) before the measurements.\\
\begin{figure}[htbp]
  \includegraphics[width=\linewidth]{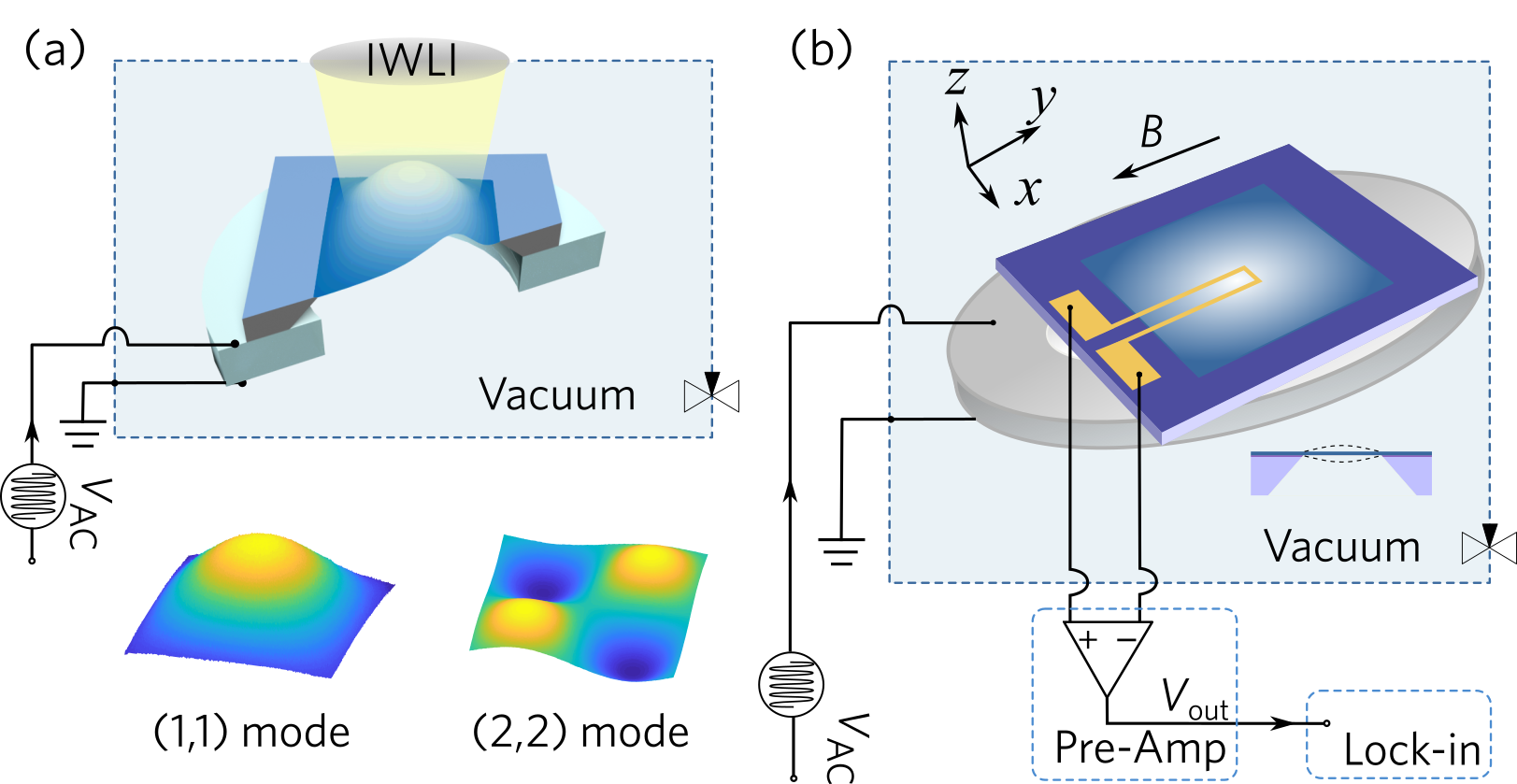}
  %\centering
  %\captionsetup{justification=centering}
  \caption{ (a) The objective of the imaging white light interferometer (IWLI) with optical measurement. Typical mode shapes of the (1,1) and the (2,2) modes are plotted below, measured for sample OM1. (b) the on-chip magnetic induced detection scheme with 27 nm thick nanoelectrode probe on top of the membrane.}
  \label{fig:setupandmembrane}
\end{figure}

\indent The chip carrying the membrane is glued to a piezo actuation ring and fixed into a vacuum chamber. Bending waves of the membranes are excited by applying an AC voltage $V_\textrm{AC}$ to the piezo. 
The cross section of the structure is presented in Fig. \ref{fig:setupandmembrane}(a). The spatial deflection patterns of the (1,1) and the (2,2) mode are demonstrated. The frequency response of the membranes can be captured by either optically  by Vibrometry In Continuous Light (VICL) in the IWLI, or inductively using the on-chip nanoelectrode probe system, shown in Fig. \ref{fig:setupandmembrane}(a) and (b). Three membrane samples OM1, EM1, and PM1 are used in this work, more details of samples and experimental systems are described in the Supplemental Material \cite{SM} similar systems were used in previous works \cite{waitz2012mode, waitz2015spatially, zhang2015vibrational, yang2019spatial, yang2021persistent, yang2021mechanically}.\\

\begin{figure}
   \includegraphics[width=\linewidth]{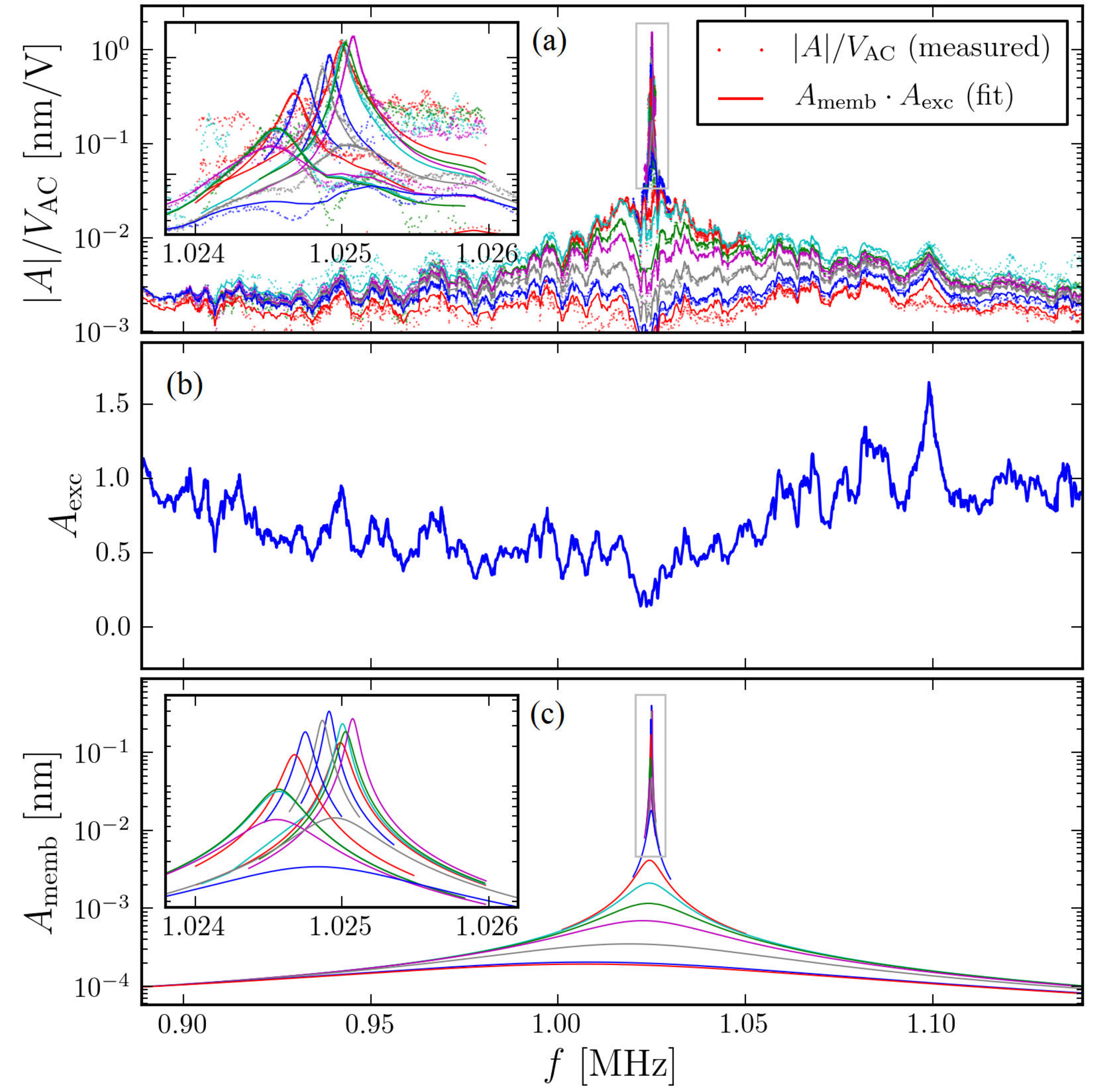}
   \centering
   \caption[s]{ The measurement and free-fitting procedure applied to sample OM1. (a) Measured resonance curves under different gas pressure are shown as dots. The gas pressure ranges from 1.6 $\times$ 10$^{-2}$ to 992 mbar, and $N = 24$ here. The solid lines are results of a fit of $|A|_j(\omega)$. (b) Amplitude of the system fluctuation $A_\textrm{exc}$ (defined in the supporting information). The product $A_\textrm{memb} \cdot A_\textrm{exc}$ (solid lines in (a)) is used to fit the experimental data.  (c) Resonance amplitudes $A_\textrm{memb}$ of membrane resonator. The inset shows a magnified view of the low pressure peaks in the light gray rectangle in the main panel.
   }
   \label{fig:fit_3N}
 \end{figure}

\begin{figure}
   \includegraphics[width=0.8\linewidth]{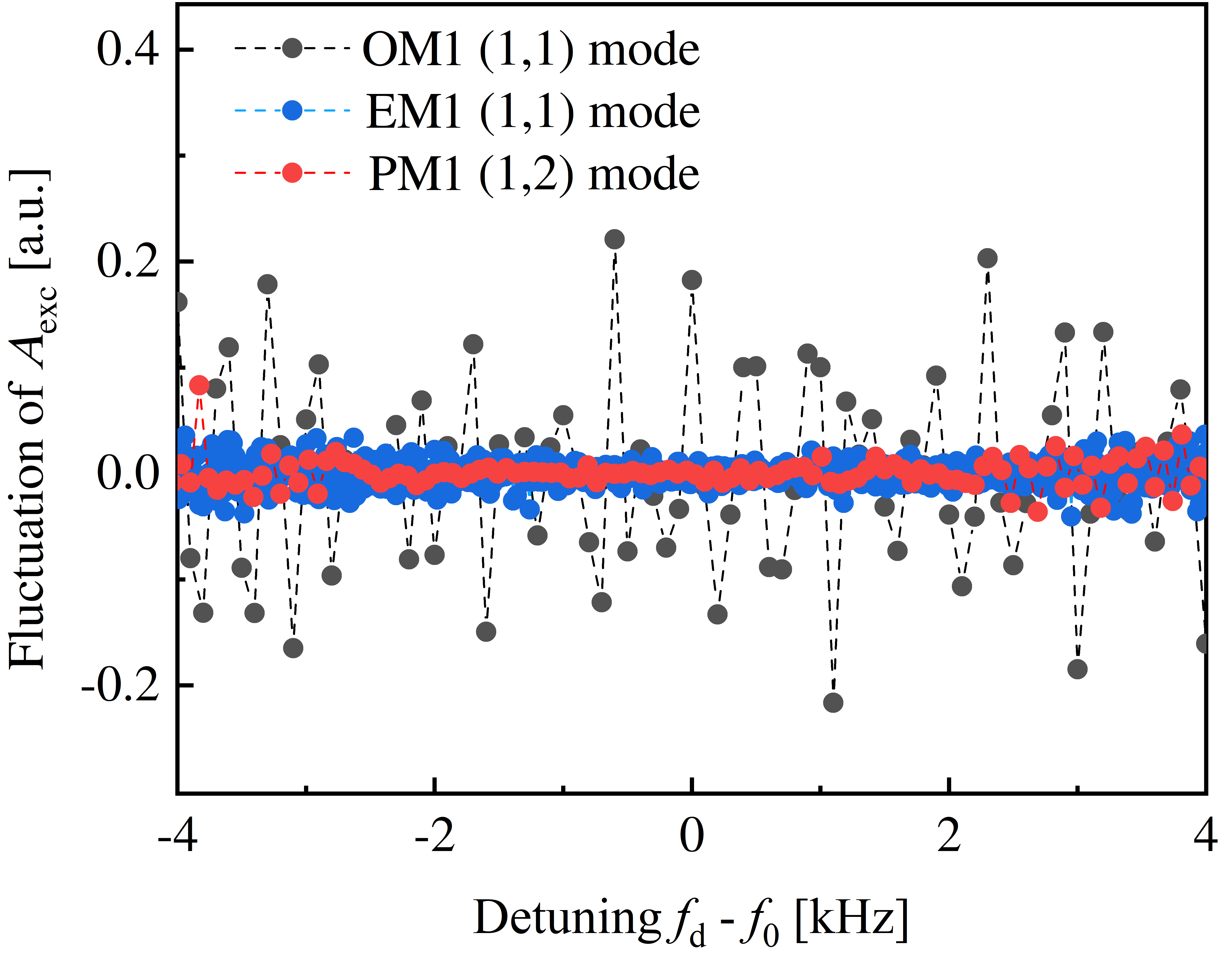}
   \caption{Comparison of the noise level of different systems. The color of the dots corresponds to different samples and their measurement system. Gray: sample OM1 in the optical measurement system, blue: sample EM1 in the electromagnetic measurement system, red: sample PM1 in the permanent-magnet measurement system. }
   \label{fig:Aexc_noise}
 \end{figure}

\begin{figure}
   \includegraphics[width=\linewidth]{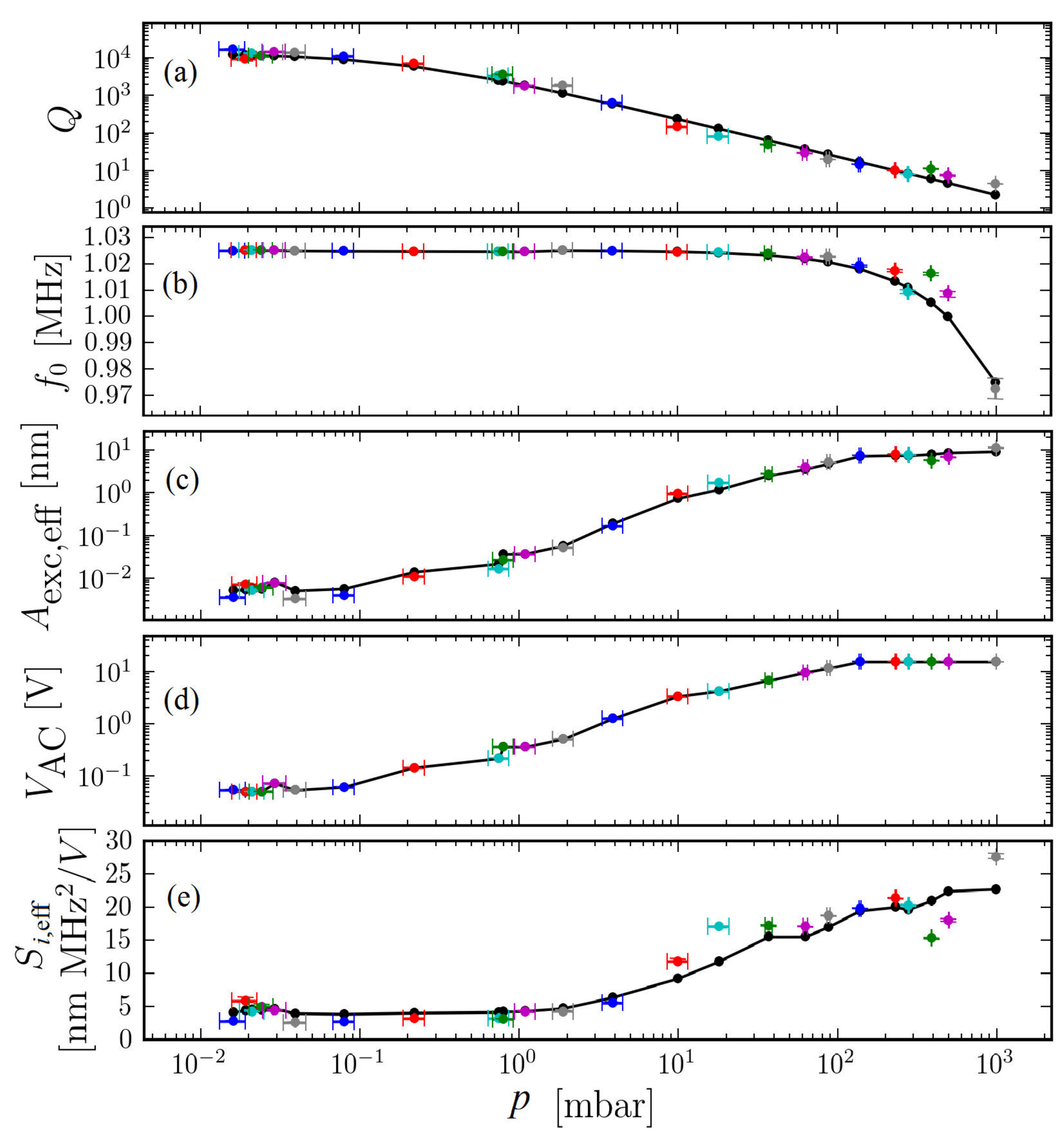}
   %\captionsetup{justification=justified}
   \caption{Visualization the fit parameters of sample OM1 as a function of the pressure. The colored dots show the results of a free fit with 3$\textit{N}$ free parameters (same color code as  in Fig. \ref{fig:fit_3N}. The black dots and lines correspond to a fit with additional constraints, see text. The quality factor $Q$ is plotted in (a), the eigenfrequency of the corresponding undamped oscillator $f_0 = \omega_0/2\pi$ in panel (b), $A_\textrm{exc,eff}$ in (c), and the excitation voltage $V_\textrm{AC}$ in (d). (e) The proportionality factors $S_{i,\textrm{eff}}$ \cite{SM} correspond to the coupling strength between the excitation oscillation and the membrane oscillation. }
   \label{fig:fit_params}
 \end{figure}

%
%
%------------------------Constrained fitting method--------------------
%
%
\indent The measured membranes are placed in an adjustable vacuum environment which acts as a well-controlled control scheme to tune the mechanical properties, especially the damping factor of the membranes, by viscous friction between membrane and the surrounding gas (see Fig. \ref{fig:setupandmembrane}). The pressure of the chamber ranges from $p = 0.001$~mbar to atmospheric pressure, for more details see SM \cite{SM}. \\
\indent The Fig. \ref {fig:fit_3N} (a) shows the different resonance curves of membrane resonators measured at different pressures. The resonance width decreases markedly as the pressure becomes lower, indicating the elimination of gas friction as a damping source. For low pressure the amplitude $|A|$ features a pronounced resonance around $f = 1.023$ MHz superimposed by small amplitude fluctuations throughout the whole frequency range. However, at higher pressure the amplitude of the membrane resonator decreases to the background fluctuation floor, thereby  hiding the shape of the resonance curve.  \\
\indent To disentangle the intrinsic resonance response of the membrane from system fluctuations, we develop a free-fitting method to extract the fluctuations contributed by the drive and measurement system from multiple damping-controlled resonance curves. In the present work, we do not aim at understanding the mechanical details of the highly complex drive system consisting of the piezo, the supporting metal block, the chip as well as the glue films in between, nor the differences between different detection methods. In the following, we therefore refer to the impact of the instrumentation and the environment as to the ``excitation system'' and denote its resonance amplitude as $A_\textrm{exc}(\omega)$.\\
\indent For a linear excitation system, the oscillation amplitude of the mechanical support of the membrane is a product of a function $A_\textrm{exc}(\omega)$ and of $V_\textrm{AC}$, which is a pure function of $\omega$. The measured amplitude $|A|$ in the linear regime is a product of the resonance $A_\textrm{memb}(\omega)$ of an isolated membrane and the oscillation amplitude of the excitation system
\begin{equation}
 |A| = A_\textrm{memb}(\omega) \cdot A_\textrm{exc}(\omega) \cdot V_\textrm{AC}\;. \label{AAAV}
\end{equation}
For $A_\textrm{memb}$ the well-known resonance curve of a damped driven harmonic oscillator is assumed
\begin{equation}
 A_\textrm{memb} = \frac S {\sqrt{(\omega^2-\omega_0^2)^2 + 4\beta^2\omega^2}} \;,\label{Amemb}
\end{equation}
with an eigenfrequency $\omega_0$, a damping constant $\beta$ and a proportionality factor $S$ measuring the coupling strength between the membrane oscillation and the excitation oscillation.\\
\indent Using this procedure, $|A|$ and $V_\textrm{AC}$ can be obtained from the experiment. As shown in Fig. \ref{fig:fit_3N} (a), the $|A|$ is characterized by the optical measurement system as a function of $\omega$ for $N$ different pressures $p_i$ of the surrounding atmosphere, and $|A|_i$ is corresponding to pressure $p_i$. The parameters  $\omega_{0i}$, $\beta_i$ and $S_i$ in Eq. (\ref{Amemb}) for $A_\textrm{memb}$ as well as the excitation voltage $V_{\textrm{AC}i}$ are also allowed to be pressure dependent. If all parameters are given, $A_\textrm{exc}(\omega)$ can be calculated:
\begin{align}
 &A_\textrm{exc}(\omega,\{\omega_{0i}\},\{\beta_i\},\{S_i\}) \notag\\
 &\qquad = \sum_{i=1}^N W_i(\omega,\{\omega_{0i}\},\{\beta_i\},\{S_i\})\notag\\
 &\qquad \qquad \cdot \frac{|A|_i(\omega)}{A_\textrm{memb}(\omega,\omega_{0i},\beta_i,S_i) V_{\textrm{AC}i}} \;.\label{Aexc}
\end{align}
According to Eq. (\ref{AAAV}) the fraction on the bottom right is equal to $A_\textrm{exc}(\omega)$ for each addend. Therefore the weighted average using a weight function with $\sum_i W_i \equiv 1$ is also equal to $A_\textrm{exc}(\omega)$.
The best fit values for $\{\omega_{0i}\}$, $\{\beta_i\}$ and $\{S_i\}$ can then be obtained:
\begin{align}
 \forall j: \quad |A|_j(\omega) = &A_\textrm{memb}(\omega,\omega_{0j},\beta_j,S_j) \label{fit_eq}\\
 &\cdot A_\textrm{exc,0}(\omega,\{\omega_{0i}\},\{\beta_i\},\{S_i\}) \cdot V_{\textrm{AC}j}\notag\\
 \textrm{with} \qquad &A_\textrm{exc,0}(\omega,...)=\frac {A_\textrm{exc}(\omega,...)}{\langle A_\textrm{exc}(\omega',...) \rangle_{\omega'}}  \notag
\end{align}
where $A_\textrm{exc,0}$ is $A_\textrm{exc}$ normalized by its respective pressure dependent frequency average, see SM \cite{SM} for the detailed explanation of the procedure. \\
\indent In Fig. \ref{fig:fit_3N}, the factors $A_\textrm{memb}(\omega)$ and $A_\textrm{exc}(\omega)$ of Eq. (\ref{AAAV}) deduced by this fitting method are plotted as a function of the excitation frequency. The dimensionless response of the excitation system, $A_\textrm{exc}$, shown in Fig. \ref{fig:fit_3N} (b) fluctuates around an average of 0.55, but has no pronounced feature in the relevant frequency range. Fig. \ref{fig:fit_3N} (c) displays the extracted intrinsic response of the membrane $A_\textrm{memb}$, where now the resonance is clearly discernible also for high pressures.\\
\indent To verify the universality of this noise extraction method in mechanical systems, we also introduce an on-chip magnetic inductive detection for membrane resonators (Fig. \ref{fig:setupandmembrane} (b)) with different size and structure, samples EM1 and PM1. The detailed data processing for the noise extraction and the obtained discernible resonance curves of samples EM1 and PM1 under high-pressure atmosphere can be found in the SM \cite{SM}.  The extracted fluctuations using these three different measurement systems are compared in Fig. \ref{fig:Aexc_noise}. \\
\indent As shown in Fig. \ref{fig:Aexc_noise}, the extracted fluctuations of $A_\textrm{exc}$ in the optical measurement are more pronounced than that of the inductive measurements. Since all three membrane resonators in Fig. \ref{fig:Aexc_noise} are mounted to the same type of piezo with similar parameters, similar contributions of the fluctuations from the drive system are expected to be present. However, the main noise sources of the inductive and optical measurements are different. In the inductive measurements, the main noise source, especially at high pressure, has been found to be the cross-talk of the piezo and its wiring in the chamber, yet the common-mode noise (including the vibration of the platform, random noise from the environment, etc.) has been cancelled by the differential amplification in the detection circuit. Therefore, the contribution of the fluctuations from the inductive measurement can be expected to be minimized when sweeping the drive frequency. In contrast, the main noise source of the optical measurement arises from the mechanical vibration of the active vibration isolation damping table under the sample chamber. This mechanical vibration of the table is random and at low frequency. Its influence on the mechanical motion of the membrane resonator leads to random fluctuations in a wide frequency range. As shown in Fig. \ref{fig:Aexc_noise}, these differences between the inductive and optical measurement have been successfully characterized and directly observed from the extracted fluctuation curves of the response curves by using the free-fitting procedure.\\
\indent In the procedure to separate the fluctuations of the excitation system and the intrinsic amplitude of membrane resonator, the mechanical parameters of membrane resonator can also be well-fitted. Figure \ref{fig:fit_params} shows the best fit parameters ($3N$ free parameters: $\{\omega_{0i}\}$, $\{\beta_i\}$ and $\{S_i\}$) of sample OM1 as a function of pressure as colored dots. The quality factor $Q=\omega_0/\beta$ (see Fig. \ref{fig:fit_params} (a)) increases by almost 4 orders of magnitude when the pressure decreases down to 50 mbar. At low pressure ($p < 0.05$ mbar), the $Q$ factor saturates, indicating that the damping is dominated by intrinsic properties such as clamping losses. In Fig. \ref{fig:fit_params} (b), the eigenfrequency $f_0 = \omega_0/2\pi$ is shown which is almost independent of the pressure up to values of $p \simeq 30$ mbar. The decrease at higher pressure indicates that the strong damping limit is approached.  The effective amplitude of the excitation system with $A_{\textrm{exc,eff},i} = S_{i}A_\textrm{exc}(\omega)V_\textrm{AC}/\omega_0^2$ is plotted in panel (c). It is proportional to the excitation voltage $V_\textrm{AC}$ shown in panel (d).  The effective amplitude factors $S_{i,\textrm{eff}} = S_{i}A_\textrm{exc}(\omega)$ are shown in panel (e), more details see SM \cite{SM}.\\ 
%
%
%------------------------Constrained fitting method--------------------
%
%
Based on the free-fitting method with $3N$ parameters, we further develop a constrained fitting method to test physical models for the environmental influences, expressed as  mathematical relations. %In this method, the constraints are assumptions that create mathematical links among parameters which have specific physical meanings in the mechanical system. 
By introducing these additional constraints into the free-fitting procedure, the number of independent fitting parameters is largely reduced.

\begin{figure}
   \includegraphics[width=0.9\linewidth]{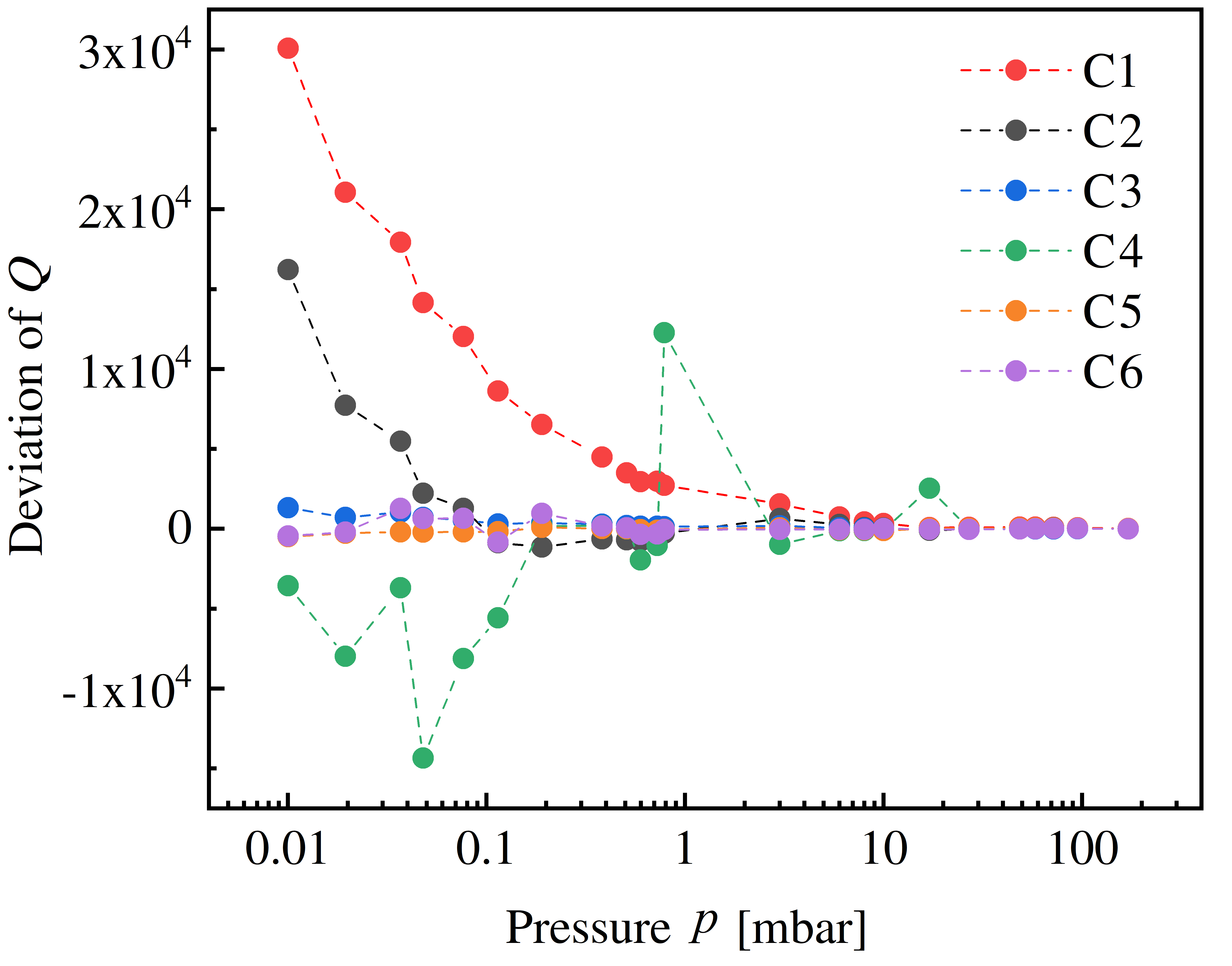}
   \caption{Comparison of different constrained fitting models as a function of the pressure for resonator EM1. The colored dots show the difference between the result of the free-parameter fitting and constrained fittings.}
   \label{fig:constrain_fitting}
 \end{figure}

Here, we provide the following six, physically motivated constraints which might exist in mechanical resonators (motivation and mathematical expressions are given in the SM \cite{SM}):\\
(1) $A_\textrm{exc,0}$ is drive-amplitude dependent. The correction is approximated by a power law of 2nd order:   $1+bA_\textrm{memb}V_\textrm{AC}+a(A_\textrm{memb}V_\textrm{AC})^2$.\\
(2) $A_\textrm{exc,0}$ is no longer a global variable extracted by fitting all curves, but 
within each decade of $V_{\textrm{AC}}$.\\
(3) The $S_i$ are fixed in each decade of $V_\textrm{AC}$ but are allowed to vary from decade to decade.\\
(4) The $S_i$ are excitation dependent following a power law: $S_i = S_{i,0} + S_{i,1}\cdot V_\textrm{AC} + S_{i,2}\cdot V_\textrm{AC}^2 + S_{i,3}\cdot V_\textrm{AC}^3$.\\
(5) $f_0$ depends linearly on the pressure for $p \geq$ 1.5 mbar.\\
(6) $\beta$ depends linearly on the pressure.\\
\indent We use three of them (3), (5), and (6) for the constrained fitting process shown in Fig. \ref{fig:fit_params}. The results are plotted as black dots and lines, with the number of independent fitting parameters ranging from $3N = 72$ to $N-5 = 19$, for details see SM \cite{SM}. We obtain a good agreement between the free-fitting results and the constrained model. \\
\indent Furthermore, different combinations of constraints can be flexibly chosen and tested for different mechanical systems. Fig. \ref{fig:constrain_fitting} shows the deviations between the $Q$ factor of resonator EM1 obtained from the free-fitting method and the constrained fitting method. The combinations of constraints are C1: (1, 3, 5, 6); C2: (2, 3, 5, 6); C3: (1, 3, 5); C4: (1, 4, 5); C5: (1, 5); C6: (2). As shown in Fig. \ref{fig:constrain_fitting}, C3, C5, C6 show better consistency with the results of the free fitting; an enlarged plotting of the remaining systematic deviations for these three models are shown and discussed in the SM \cite{SM}. The most important conclusions are that for C3 the deviation of the $Q$ factor shows a weak decaying tendency, while for C5 the deviation of $Q$ increases with the pressure. The main difference is introduced by the constraint (3).  C6 individually tested the assumption (2); the deviation fluctuates with small amplitude non-systematically around 0. The influence of individual assumptions as a constraint to the system is discussed in the SM \cite{SM}. By decomposing the combined constraints into the individual constraints, we clearly demonstrate that the constraints (4) and (6) are inappropriate assumptions for the given system and lead to major deviation in the combined constraints C1, C2 and C4. \\
\indent Both the free and the constrained fitting procedures 
%is important for extreme sensitive mechanical systems, and 
can in principle be extended to all resonating systems in different physical fields, e.g., in spectroscopy, optomechanics, spin electronics, quantum mechanics and are applicable for extreme sensitive systems. Once a function of a specific physical model such as defects, crystallization or temperature-stress relation of the resonator has been established, it can be utilized as a constraint. In the SM we also show how the method can be applied do detect tiny temperature-induced shifts of the eigenfrequency. The validity of such a constraint can be verified by comparing the variance of the fitting results between the free and the constrained fitting procedure. When constraints describe the physical system inaccurately, the fitting results will deviate systematically from the free-fitting model.\\
%
%------------------------Conclusion--------------------
%
\indent In conclusion, we developed a free-fitting method to separate the fluctuations from the measured frequency response curves and to thereby obtain the intrinsic mechanical vibration amplitude of tunable mechanical systems in the dynamic range. By testing the free-fitting method on different samples in two measurement schemes, we show that even for small signal-to-noise ratio of the response it is possible to discern the intrinsic frequency response of the measured mechanical resonators and quantitatively extract their mechanical parameters such as the quality factor, the eigenfrequency, and the drive coupling strength. Furthermore, we also demonstrate how to set constraints by imposing mathematical links between different mechanical parameters and system variables into the fitting to test the validity of physical assumptions. In our realization, through the constrained fitting procedure, we reveal that the excitation system can either be considered to give a pressure-independent response or its influence can be described by a parabolic relation with the vibration amplitude of membrane resonators. Furthermore we show that the eigenfrequency depends linearly on the pressure in the low vacuum range. Both fitting methods have no strict prerequisites and thus have great potential to be applied in various fields of physics and engineering.\\
%
%------------------------acknowledgement--------------------
%

\indent The authors thank S. N{\"o}{\ss}ner and M. Hertkorn for their contribution in the early phase of this project. We are indebted to M. Hettich, G. Rastelli, Y. Jiang, and the SFB1432 NEMS Discussion Group for fruitful discussion and comments about the manuscript. The authors thank the Deutsche Forschungsgemeinschaft for financial support through SFB767 (project number 32152442) and SFB1432 (project number 425217212). FY appreciates financial support by the China Scholarship Council and MF by the Alexander von Humboldt Foundation.
%
%------------------------references--------------------
%
\bibliographystyle{apsrev4-1}
\bibliography{refs_air}

\end{document}

\title{Supplemental materials for the manuscript of ``Quantitative signal extraction in the dynamic range of nanomechanical systems by free and constrained fitting''}
\author{Fan Yang}
\affiliation{Department of Physics, Universit{\"a}t Konstanz, 78464 Konstanz, Germany}
\author{Reimar Waitz}
\altaffiliation[Now at: ]{RATIONAL AG, Iglinger Str. 62, 86899 Landsberg am Lech, Germany}
\author{Mengqi Fu}
 \email{mengqi.fu@uni-konstanz.de}
\affiliation{Department of Physics, Universit{\"a}t Konstanz, 78464 Konstanz, Germany}
\author{Elke Scheer}
\affiliation{Department of Physics, Universit{\"a}t Konstanz, 78464 Konstanz, Germany}

\maketitle

\section{Sample and experimental setups}

\indent The SiN membranes are fabricated from a 0.5 mm thick commercial (100) silicon wafer. Both sides of the silicon substrate are coated with $\sim$ 500 nm thick low-pressure chemical vapor deposited (LPCVD) SiN. The membrane is fabricated on the front layer. The backside layer serves as an etch mask. Laser ablation is used to open the etch mask with a typical size of 1 $\mathrm{\times}$ 1 mm$^{2}$. Using anisotropic etching in aqueous potassium hydroxide (KOH), a hole is etched through the openings of the mask. After the KOH solution reaches the topside layer, a membrane is formed, supported by a massive silicon frame. In the present work, the membrane is  $\sim$ 500 nm thick and 
several hundreds $\mu$m in lateral scale. The chip carrying the membrane is glued to a piezo ring of 20 mm diameter and 5 mm thickness using a two-component adhesive.\\

\noindent Sample details used in the maintext:\\
OM1: SiN membrane with 400 nm thick and lateral size of 416 $\times$ 398 $\mu$m$^2$.\\
EM1: SiN membrane with 500 nm thick and lateral size of 542 $\times$ 524 $\mu$m$^2$.\\
PM1: SiN+$\textrm{SiO}_2$ double layer membrane with consisting of 125 nm SiN layer and 345 nm $\textrm{SiO}_2$. The lateral size of the membrane is 766 $\times$ 734 $\mu$m$^2$.\\
\indent In this work, we studied samples using two different measurement principles, realized in three experimental setups:\\
\indent (1) Sample OM1 has been studied by an optical method: The surface of the membrane is observed by an imaging white light interferometer (IWLI) using different light sources, described in detail in Ref. \cite{petitgrand20013d}. We utilized continuous light for recording the resonance curves, stroboscopic light was utilized to measure the deflection patterns \cite{yang2019spatial, yang2021persistent, zhang2015vibrational} using an IWLI. The sample is placed in a vacuum chamber; the chamber can be placed on top of a temperature controllable hot plate for temperature dependent measurements, the chamber connected to a pressure controller, to ensure full control over the pressure of the surrounding atmosphere in a pressure range from $P$ = 0.001 mbar to atmospheric pressure for the measurement system. For the SiN membrane resonators here, for pressures below 10$^{-2}$ mbar, the damping is dominated by intrinsic damping mechanisms of the membrane an its clamping, while losses due to the coupling to the atmosphere are negligible. After installing and pumping the sample, the system is allowed to achieve thermal equilibrium. The temperature is monitored by a thermometer attached to the sample holder in close vicinity to the substrate.  The excitation voltage is applied using a sinusoidal function generator the phase of which can be locked to the stroboscopic light of the IWLI. For measuring resonance curves, it is necessary to adjust the amplitude of the excitation $V_\textrm{AC}$ when varying the frequency to keep the vibration in the linear state within the dynamic range. \\
\indent (2) Sample EM1 and PM1 are studied by an inductive method: We use the magnetic induction method to characterize the amplitude of the membrane \cite{yang2021mechanically}. As described in the main text  and Fig. 1 (b), the whole device is placed in the vacuum chamber with a adjustable pressure at room temperature and subject to an in-plane magnetic field $B$. The magnetic field can be provided by an electromagnet or by a permanent magnet. A $\sim$ 27 nm thick Al electrode is deposited onto the membrane to form detection electrodes perpendicular to the magnetic field and their peripheral leads in parallel to the magnetic field. For the sample used in this manuscript, the detection electrode oriented perpendicular to the magnetic field is located at the center of the membrane and has a length of $L=30~\mu$m in order to detect the maximal vibration amplitude of the fundamental mode. When the membrane vibrates, the magnetic flux through the area enclosed by the detection electrode and peripheral leads changes, and thus a potential difference is generated across the Al structures on the membrane. The generated potential difference is first fed to the two input ports of a differential preamplifier to be converted into a single-ended output voltage and to be amplified by a factor ($G_\textrm{diff}$) of 500. Then the output voltage ($V_\textrm{out}$) is further measured by a lock-in amplifier. The usage of a differential preamplifier can suppress the common-mode noise (such as the noise generated in the wires and from the vibration of the sample stage) efficiently. Note that the vibration of the membrane parts under the peripheral leads does not contribute to the $V_\textrm{out}$ because the peripheral leads are parallel to the magnetic field. Therefore, the vibration velocity ($v$) of the membrane parts under the detection electrode is linearly related to the $V_\textrm{out}$ by a factor of $1/(G_\textrm{diff}BL)$, where $B=0.45$ T for electromagnet and 0.50 T for permanent magnet. Hence, when the membrane is driven by the piezo with the drive voltage of $V_\textrm{AC}$ at the frequency of $\omega_{d}=2\pi f_{d}$, the velocity is $v(t) = Aw_\textrm{d}\cos(w_\textrm{d}t)$ and the real vibration amplitude ($A$) at the position of the detection lead can be easily calculated by
\begin{equation}
    A = \frac{V_\textrm{out}}{BL\omega_{d}G_\textrm{diff}} ~.
    \label{eq:nm_conver}
\end{equation}
%

%%%%%%%%%%%%%%%%%%%%%%%%%%%%%%%%%%%%%%%%%%%%%%%%%%
%
%
\section{Free-fitting procedure}
\subsection{Free-fitting model}
\noindent The properties of the ``excitation system'' $A_\textrm{exc}(\omega)$ change from sample to sample, because of their dependence on the coupling strength of the sample to the excitation system. Here we will present how they can be determined from the total response of the system. Three preconditions have to be fulfilled for the method to be applicable: (1) The amplitude of the excitation system shows a linear dependence on $V_\textrm{AC}$. This requirement is fulfilled for small and moderate $V_\textrm{AC}$. (2) Effects of the surrounding atmosphere on the excitation system are negligible. This assumption is justified because of its high mass and low surface area. (3) Feedback by the membrane motion on the excitation system is negligible. This is fulfilled for a sufficiently large mass ratio between the excitation system and the membrane. In our case it amounts to five orders of magnitude.\\
\indent As a consequence of linearity, the oscillation amplitude of the mechanical support of the membrane is a product of a function $A_\textrm{exc}(\omega)$ and $V_\textrm{AC}$. Since the excitation system resonance $A_\textrm{exc}(\omega)$ is assumed not to depend on properties of the atmosphere or of the membrane, mentioned as precondition (2) and (3), it is a pure function of $\omega$.\\
\indent If the amplitudes are sufficiently small, the membrane can be modeled as a linear oscillator. Therefore the total absolute amplitude $|A|$ of the membrane oscillation is a product of the resonance $A_\textrm{memb}(\omega)$ of an isolated membrane and the oscillation amplitude of the excitation system
\begin{equation}
 |A| = A_\textrm{memb}(\omega) \cdot A_\textrm{exc}(\omega) \cdot V_\textrm{AC}\;. \label{AAAV}
\end{equation}
\indent In Eq. (\ref{AAAV}), only $|A|$ and $V_\textrm{AC}$ are known from the experiment. For $A_\textrm{memb}$ the well known resonance curve of a damped driven harmonic oscillator is assumed
\begin{equation}
 A_\textrm{memb} = \frac S {\sqrt{(\omega^2-\omega_0^2)^2 + 4\beta^2\omega^2}} \;,\label{Amemb}
\end{equation}
with an eigenfrequency $\omega_0$, a damping constant $\beta$ and a proportionality factor $S$ measuring the coupling strength between the membrane oscillation and the excitation oscillation provided by the piezo ring. In the following, we present a way to separate the membrane resonance $A_\textrm{memb}$ from the excitation system resonance $A_\textrm{exc}$ and to obtain the parameters $\omega_0$, $\beta$ and $S$.\\
\indent The amplitude $|A|$ is measured as a function of $\omega$ for $N$ different pressures $p_i$ of the surrounding atmosphere, and $|A|_i$ is the corresponding amplitude. The parameters  $\omega_{0i}$, $\beta_i$ and $S_i$ in Eq. (\ref{Amemb}) for $A_\textrm{memb}$ as well as the excitation voltage $V_{\textrm{AC}i}$ are also allowed to be pressure dependent. If all the parameters are given, $A_\textrm{exc}(\omega)$ can be calculated:
\begin{align}
 &A_\textrm{exc}(\omega,\{\omega_{0i}\},\{\beta_i\},\{S_i\}) \notag\\
 &\qquad = \sum_{i=1}^N W_i(\omega,\{\omega_{0i}\},\{\beta_i\},\{S_i\})\notag\\
 &\qquad \qquad \cdot \frac{|A|_i(\omega)}{A_\textrm{memb}(\omega,\omega_{0i},\beta_i,S_i) V_{\textrm{AC}i}} \;.\label{Aexc}
\end{align}\\
\indent According to Eq. (\ref{AAAV}) the fraction on the bottom right is equal to $A_\textrm{exc}(\omega)$ for each addend. Therefore the weighted average using a weight function with $\sum_i W_i \equiv 1$ is also equal to $A_\textrm{exc}(\omega)$. The choice of $W_i$ is arbitrary. We use
\begin{align*}
 W_i(\omega,\{\omega_{0i}\},\{\beta_i\},\{S_i\}) = \frac{w_i(\omega,\omega_{0i},\beta_i,S_i)}{\sum_j w_j(\omega,\omega_{0j},\beta_j,S_j)} \\
 w_i(\omega,\omega_{0i},\beta_i,S_i) = \frac{A_\textrm{memb}(\omega,\omega_{0i},\beta_i,S_i)}{\max_{\omega'}[A_\textrm{memb}(\omega',\omega_{0i},\beta_i,S_i)]}
\end{align*}
with $\max_\omega[A_\textrm{memb}]$ denoting the maximum of the membrane resonance curve. This choice of the weight function is advantageous, because data points with high amplitude and therefore higher signal-to-noise ratio (SNR) are given a higher weight. In Fig. 2 the factors of Eq. (\ref{AAAV}) deduced by this analysis are plotted as a function of the excitation frequency. The dimensionless response of the excitation system, $A_\textrm{exc}$, shown in panel (b) fluctuates around an average of 0.55, but has no pronounced feature in the relevant frequency range. Panel (c) displays the corrected response of the membrane $A_\textrm{memb}$, where now the resonance is clearly discernible also for high pressures.\\
In Eq. (\ref{AAAV}) $A_\textrm{memb}$ and $A_\textrm{exc}$ depend on the parameters $\{\omega_{0i}\}$, $\{\beta_i\}$ and $\{S_i\}$. Therefore it can be used to find the best fit values for $\{\omega_{0i}\}$, $\{\beta_i\}$ and $\{S_i\}$:
\begin{align}
 \forall j: \quad |A|_j(\omega) = &A_\textrm{memb}(\omega,\omega_{0j},\beta_j,S_j) \label{fit_eq}\\
 &\cdot A_\textrm{exc,0}(\omega,\{\omega_{0i}\},\{\beta_i\},\{S_i\}) \cdot V_{\textrm{AC}j}\notag\\
 \textrm{with} \qquad &A_\textrm{exc,0}(\omega,...)=\frac {A_\textrm{exc}(\omega,...)}{\langle A_\textrm{exc}(\omega',...) \rangle_{\omega'}}  \notag,
\end{align}
where $A_\textrm{exc,0}$ is $A_\textrm{exc}$ normalized by its respective, pressure dependent frequency average. If $(\{\omega_{0i}\},\{\beta_i\},\{S_i\})$ is a solution, without normalization, using $A_\textrm{exc}$ instead of $A_\textrm{exc,0}$ in Eq. (\ref{fit_eq}), $(\{\omega_{0i}\},\{\beta_i\},\{cS_i\})$ with an arbitrary constant $c\ne 0$ will also be a solution, since $A_\textrm{memb}$ in Eq. (\ref{Amemb}) is proportional to $c$ and $A_\textrm{exc}$ in Eq. (\ref{Aexc}) is proportional to $1/c$. This ambiguity is avoided by the normalization of $A_\textrm{exc}$ to $A_\textrm{exc,0}$.\\
\indent A least-square optimization algorithm is used to compute an approximate solution for $\{\omega_{0i}\}$, $\{\beta_i\}$ and $\{S_i\}$. The number of free parameters is $3N$, but can be reduced by forcing additional constraints on the so far independent parameters, as we will demonstrate in later discussion.\\
\indent A graph of the best fit parameters for a fit with $3N$ free parameters as a function of pressure is shown as colored dots in Fig. 4. in the main text. Instead of the damping parameter $\beta$ the equivalent but more accessible quality factor $Q=\omega_0/\beta$ is shown in the upper panel. The $Q$ factor increases by almost 4 orders of magnitude when lowering the pressure from atmospheric pressure to 50 mbar. At very low pressure ($p < 0.05$ mbar) the $Q$ factor saturates, indicating that the remaining damping is given by intrinsic properties of the sample. In the second panel, the resonance frequency of the corresponding undamped oscillator $f_0 = \omega_0/2\pi$ is shown.  $f_0$ is almost independent of pressure up to values of $p \simeq 30$ mbar. For higher pressure it decreases, indicating that the strong damping limit is approached. Since the parameters $S_i$  strongly depend on the normalization of $A_\textrm{exc,0}$ of each membrane resonance curve, and $A_\textrm{exc,0}$ a group of $N$ curves denotes which a matrix, we show instead in panel (c) of Fig. 4 the more descriptive effective excitation amplitude of the membrane frame $A_\textrm{exc,eff}$, which is defined as follows: Comparing Eq. (\ref{AAAV}) and (\ref{Amemb}) to the well known formulas of the driven damped harmonic oscillator, we conclude that the amplitude of the excitation system is $S_iA_\textrm{exc}(\omega)V_\textrm{AC}/\omega_0^2$. The effective amplitude is defined as the excitation system amplitude averaged using $A_\textrm{memb}$ as a weight function
\begin{align}
 A_{\textrm{exc,eff},i} = \frac {\int \mathrm{d}\omega [A_\textrm{memb}(\omega,...) \cdot \frac{S_iA_\textrm{exc}(\omega)V_\textrm{AC}}{\omega_0^2}]}  {\int \mathrm{d}\omega A_\textrm{memb}(\omega,...)} \;.\label{Aexc_eff}
\end{align}
The $A_\textrm{exc,eff}$ of the excitation system has a similar pressure dependence as $V_\textrm{AC}$, as anticipated. $V_\textrm{AC}$, shown in Fig. 4 (d) was varied roughly linearly with pressure to keep the absolute amplitude within a similar range. The effective amplitude factors $S_{i,\textrm{eff}}$ in panel (e) are defined as:
\begin{align}
 S_{i,\textrm{eff}} = \frac {A_{\textrm{exc,eff},i}}  {V_\textrm{AC}} \cdot \omega_0^2\label{S1_eff} ~.
\end{align}
This first analysis shows that for a given pressure the resonance curves can successfully be decomposed into the contribution of the excitation system and the system itself, justifying the assumptions made at the beginning. However, this free fitting with $3N$ parameters does not reveal the reasons for the pressure dependence of $Q$ and $f_0$. To do so, several additional constraints that relate the fitting parameters deduced for the individual pressure values can be applied.\\

%-------------------------------------
\subsection{System noise evaluation}
\begin{figure*}
  \centering
  \includegraphics[width=\linewidth]{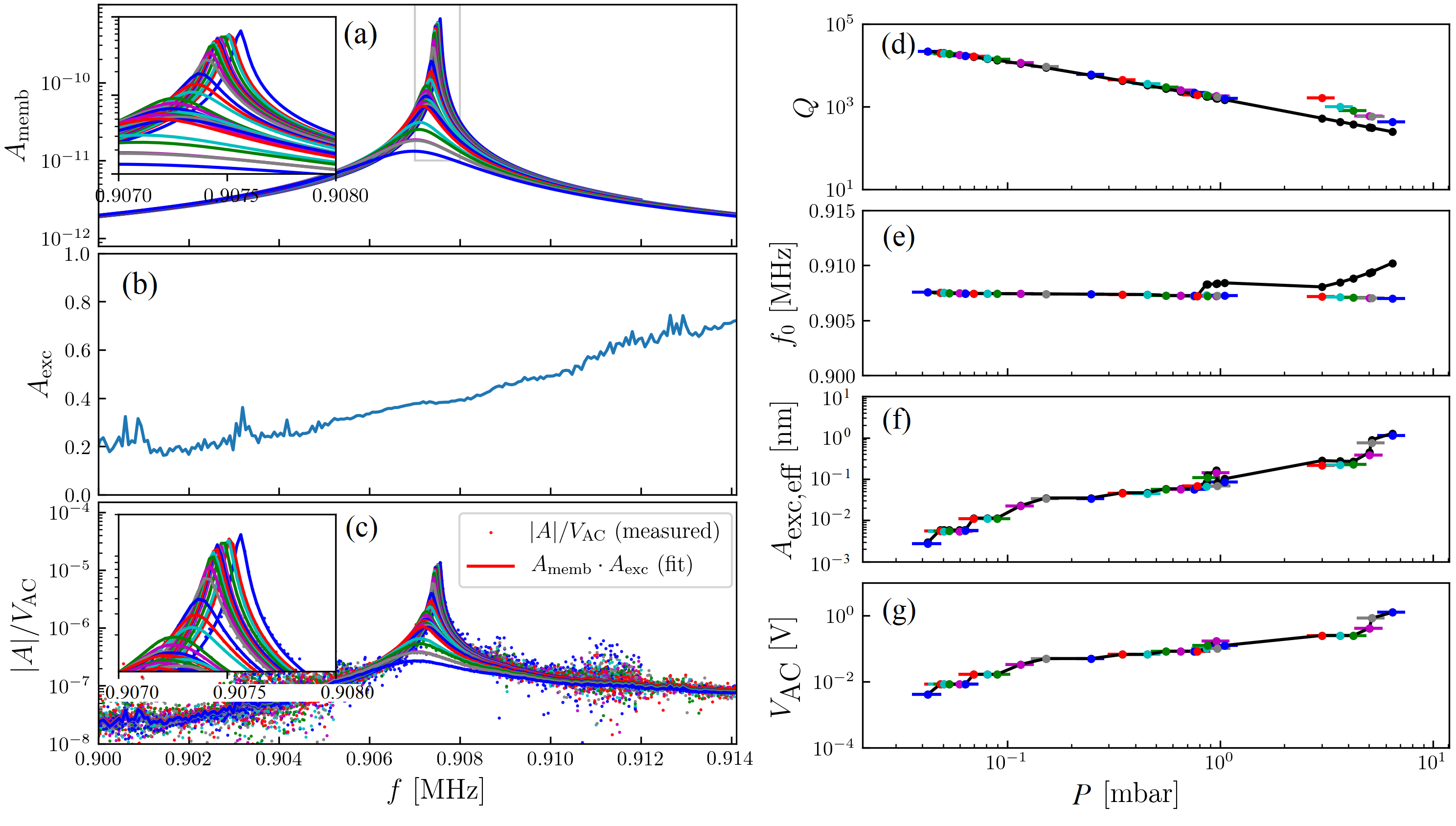}
  \caption{Results of the (1,2) mode vibration of the SiN membrane resonator PM1 around 0.907 MHz using thee free and constrained fitting method. 
  (a) The intrinsic responses of the membrane $A_\textrm{memb}$. (b) The separated $A_\textrm{exc}$. (c) The measured resonance curves under different pressures with their individual fitting. (d-g) The extracted $Q$ factor, $f_0$, effective excitation $A_{\textrm{exc,eff}}$ and excitation voltage $V_\textrm{AC}$, respectively, by free-fitting method, shown as color dots. Black dots and lines represent the corresponding constrained fitting results. }
  \label{fig:SM_PM01_combined}
\end{figure*}

\noindent As an example we discuss results obtained on sample PM1 with an on-chip nano-probe, shown in Fig. 3 of the main text. The amplitude response is captured in a magnetic field induced by a permanent magnet as a function of the pressure (180 mbar down to 0.001 mbar). The frequency range is adapted to the width of the resonance curve. The free-fitting procedure is applied to obtain the decomposed system response, i.e., the contribution of the excitation system $A_\textrm{exc}$ (shown in Fig. \ref{fig:SM_PM01_combined} (b)) and the membrane itself $A_\textrm{memb}$ (shown in Fig. \ref{fig:SM_PM01_combined} (a)). 
In Fig. \ref{fig:SM_PM01_combined} (d)-(g), the colored dots show the results of the 3$N$ free parameter fitting and using the same color code as in Fig. \ref{fig:SM_PM01_combined} (a) and (d).\\
\indent We compare three different samples in different experimental setups in order to demonstrate the capability of the free-fitting procedure to evaluate the system noise. The system noise can be extracted from the fluctuations of $A_\textrm{exc}$ by utilizing the Savitzky-Golay filter with a polynomial order of 2. The fluctuations of $A_\textrm{exc}$ for the three different systems are plotted in Fig. 3 in the main text.\\
\indent Because the main noise sources from the electromagnetic measurement systems (for example, the crosstalk from the excitation circuits) differ from the optical measurements (the vibration of the lens or substrate of the setup), the induced fluctuations vary in these measurement systems, as shown in Fig. 3 in the main text. The setups in which samples EM1 and PM1 were studied both use a differential detection scheme which efficiently reduces the common mode noise of the system compared to sample OP1 measured in the IWLI. The IWLI uses an active vibration isolation damping table. The pronounced difference in the fluctuation intensity between the optical and the inductive measurement schemes indicates that the difference arises mainly from the detection schemes.
%
%
%%%%%%%%%%%%%%%%%%%%%%%%%%%%%%%%%%%%%%%%%%%%%%%%%%
\newpage
\section{Constrained fitting procedure}
%----------------------------------------
\subsection{Combined constrained fitting}
\noindent In the constrained fitting shown in the main text, we have the following assumptions:\\
(1) $A_\textrm{exc,0}$ is drive-amplitude dependent. The correction is approximated by a power law of second order:
\begin{align}
 &A_\textrm{exc} = \sum_{i=1}^N W_i\cdot \frac{|A|_i(\omega)}{A_\textrm{memb} V_{\textrm{AC}i} \cdot (1+bA_\textrm{memb}V_{\textrm{AC}i}+aA_\textrm{memb}^2V_{\textrm{AC}i}^2)},~~ 
 A_{\textrm{exc,0}} = \frac{A_{\textrm{exc}} }{\langle A_{\textrm{exc}}(\omega',...) \rangle_{\omega'}} \;.\label{Aexc_n1}
\end{align}
(2) $A_\textrm{exc,0}$ is no longer a global variable extracted by fitting all curves, but within each decade of $V_{\textrm{AC}}$, i.e., $V_{\textrm{AC},k}$, here $k$ denotes the decade, $k \in (\lfloor \textrm{lg}(V_{\textrm{AC},i}) \rfloor - \lfloor \textrm{lg}(V_\textrm{AC,min}) \rfloor)$:
\begin{align}
 A_{\textrm{exc},k} = \sum_{i=1}^{N_k} W_{i,k}\cdot \frac{|A|_{i,k}(\omega)}{A_{\textrm{memb},k} V_{\textrm{AC}i,k}},~~ 
 A_{\textrm{exc,0},k} = \frac{A_{\textrm{exc},k} }{\langle A_{\textrm{exc},k}(\omega',...) \rangle_{\omega'}} \;.\label{Aexc_decade}
\end{align}
(3) The $S_i$ are fixed in each decade of $V_\textrm{AC}$ but are allowed to vary from decade to decade:
\begin{align}
 S_{k} = \frac{\sum_{i=1}^{N} S_{i}}{N}, ~~ k \in (\lfloor \textrm{lg}(V_{\textrm{AC},i}) \rfloor - \lfloor \textrm{lg}(V_\textrm{AC,min}) \rfloor). 
 \label{S1_decade}
\end{align}
(4) The $S_i$ are excitation dependent following a power law with $3^{rd}$ higher order:
\begin{align}
 S_{i} = S_{i,0}\cdot(1 + S_{i,1}\cdot V_\textrm{AC} + S_{i,2}\cdot V_\textrm{AC}^2 + S_{i,3}\cdot V_\textrm{AC}^3). \label{S1_poly3_Volt}
\end{align}
(5) $f_0$ depends linearly on the pressure for $p > 1.5$ mbar:
\begin{equation}
\label{f0_partial}
 \omega_{0,i}=\left\{
 \begin{aligned}
\omega_0 + \omega_{m} \cdot p_i & , & \textrm{if}~p_i \geq 1.5 \textrm{mbar}, \\
\omega_{0,i} & , & \textrm{if}~p_i < 1.5 \textrm{mbar}.
 \end{aligned}
\right.
\end{equation}
(6) $\beta$ depends linearly on the pressure.
\begin{align}
 \label{beta_P}
 \beta_{i} = \beta_{0} + \beta_{m} \cdot p_i.
\end{align}
\indent In Fig. 4 in the main text, three exemplary physically motivated constraints (3, 5, 6) are used in the constrained fitting procedure, and are explained in detail in the following: \\
\indent (3), if the coupling between the excitation system and the membrane is independent of pressure, $S$ should be single-valued throughout the whole excitation range. We set here the slightly relaxed constraint of one $S_{1}$ per excitation voltage decade. In the given example this constraint reduces the number of independent fitting parameters from $3N = 72$ (for 24 pressure values measured) to $2N+4$, since $V_\textrm{AC}$ is varied over four orders of magnitude.  \\
\indent (5), above a certain pressure threshold (1.5 mbar in the present case) we assume a linear decrease of $f_0$, reducing the number of independent parameters by the number of measurement points above the pressure threshold, plus 1 to describe the linear decrease. In the present case to $2N-7$ = 41, since 11 data points are above the threshold. \\
\indent (6), the third constraint is that $\beta$ is a linear function of the pressure, corresponding to viscous friction, resulting of $N-5$ = 19 independent fitting parameters. \\

\begin{figure*}
  \centering
  \includegraphics[width=\linewidth]{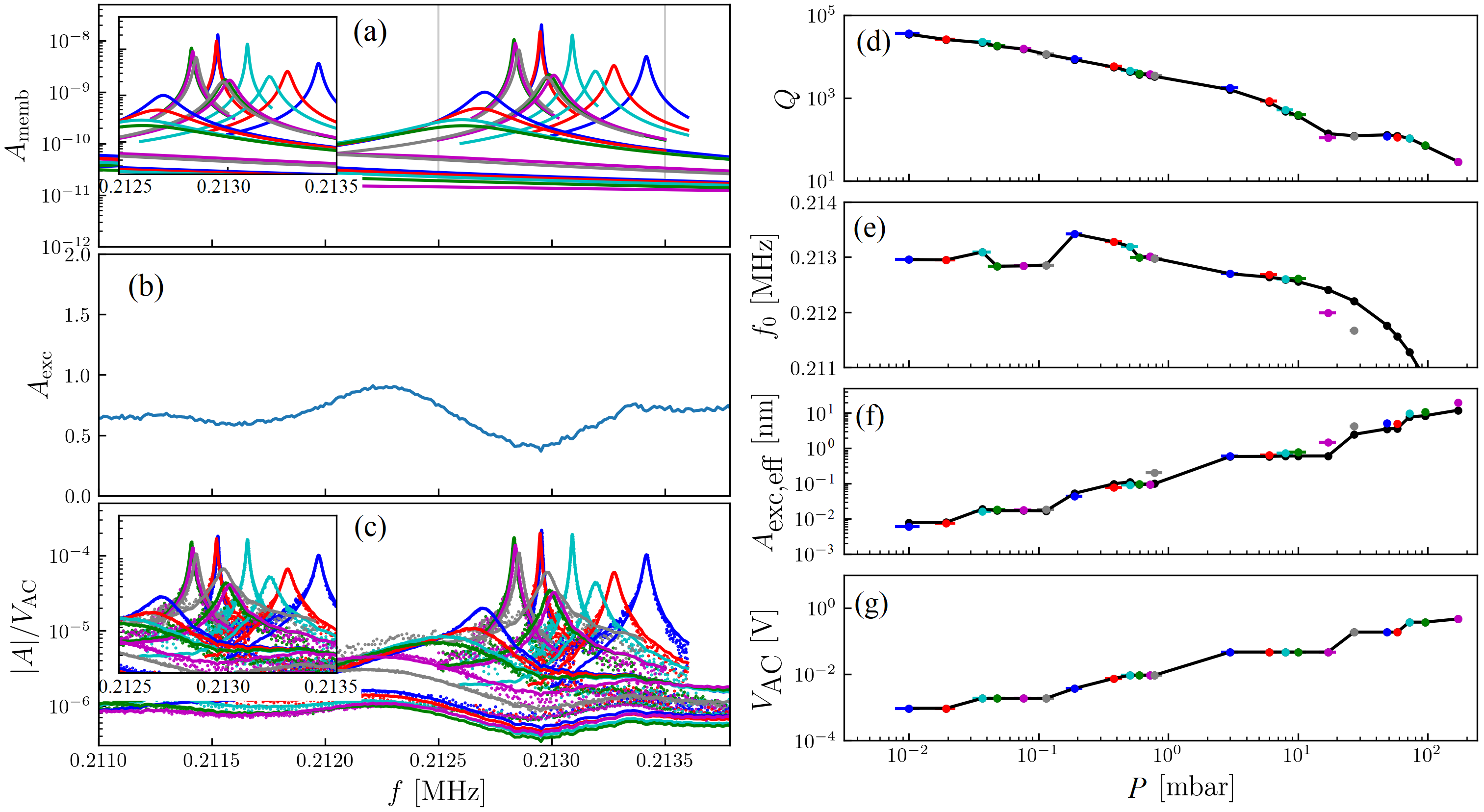}
  \caption{Result of the ground mode (1,1) vibration of sample EM1 around 0.21 MHz using thee free and constrained fitting method. 
  (a) The intrinsic responses of the membrane $A_\textrm{memb}$. (b) The separated $A_\textrm{exc}$. (c) The measured resonance curves under different pressures with their individual fitting. (d-g) The extracted $Q$ factor, $f_0$, effective excitation $A_{\textrm{exc,eff}}$ and excitation voltage $V_\textrm{AC}$, respectively, by free-fitting method, shown as colored dots. Black dots and lines represent the corresponding constrained fitting results. }
  \label{fig:SM_EM08_combined}
\end{figure*}

In Fig. 5 of the main text, the resulting deviation indicates the discrepancy between the constrained fitting results and the free fitting under this specific combination of the constraints and is a measure for the appropriateness of the used constraints. \\
\indent To generate Fig. 5 in the main text, a SiN membrane resonator (EM1) with an on-chip nanoprobe was used. The amplitude response was captured by an electromagnetic field induced signal. Same as the measurement explained in the main text, a pressure-dependent resonance test is done for pressures from 180 mbar down to 0.001 mbar, using the same color code as in Fig. \ref{fig:SM_EM08_combined}. The frequency range is individually adapted to the width of the resonance curve. The free parameter fitting procedure is applied to get the decomposed system response, i.e., the contribution of the excitation system $A_\textrm{exc}$ (shown in Fig. \ref{fig:SM_EM08_combined} (b)) and the membrane itself $A_\textrm{memb}$ (shown in Fig. \ref{fig:SM_EM08_combined} (a)). 
In Fig. \ref{fig:SM_EM08_combined} (d)-(g), the colored dots show the results of 3$N$ free parameter fitting and using the same color code as in  in Fig. \ref{fig:SM_EM08_combined} (a) and (d). 

\begin{figure*}[htbp]
  \centering
  \includegraphics[width=\linewidth]{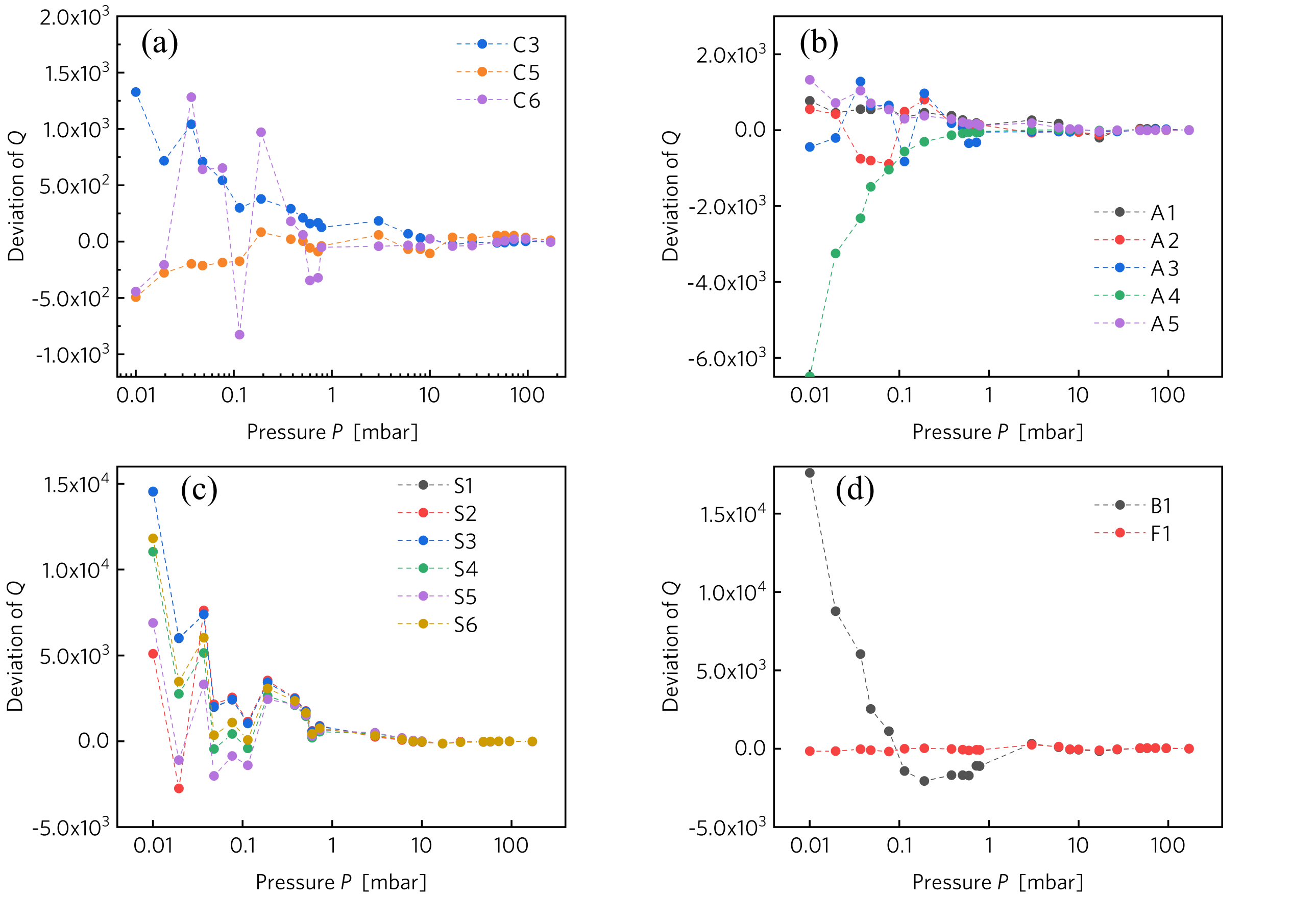}
  \caption{Constrained fitting for the (1,1) mode of sample EM1. The constraints applied in the fitting procedure are indicated in the legend (a). Blow-up of the constrained fitting results of C3, C5, and C6, the same data shown in Fig. 4 in the main text. (b) Constrained fitting results for different assumptions for $A_\textrm{exc,0}$. (c) Constrained fitting results for different assumptions for $S_\textrm{i}$.  (d) Constrained fitting results for different assumptions for $\beta$ and $f_0$.}
  \label{fig:SM_constraints_EM08}
\end{figure*}
Among the constrained fitting results shown in Fig. 5 in the main text, C3, C5, C6 show the smallest deviations from the free-fitting results. Fig. \ref{fig:SM_constraints_EM08} (a) shows a blow-up of Fig. 4 of the main text for C3, C5, and C6. The $Q$ factor fitted under the constraint C3 shows a decaying tendency, on the contrary for C5 the deviation of $Q$ increases with the pressure. The difference is introduced by the constraint (3) that  $S_i$ is kept fix within one decade of $V_\textrm{AC}$, which will be discussed later. C6 individually tested the assumption of $A_\textrm{exc,0}$ being fixed in each decade of pressure, resulting in deviations fluctuating non-systematically and with small amplitude around zero. By decomposing the combined constraints into the individual constraints, we clearly demonstrate that constraints (4) and (6) lead to  major deviation of the combined constraints C1, C2 and C4, and therefore seem to be inappropriate assumptions, as we will discuss next.\\ 
%
%
%-----------------------------------------
\subsection{Individual constrained fitting}
\indent Now we analyze each assumption as constraint individually. In Fig. \ref{fig:SM_constraints_EM08} (b), we individually tested the following assumption of the physical model for the excitation system $A_\textrm{exc,0}$. \\
A1: $A_\textrm{exc,0}$ is constant.
\begin{align}
 A_\textrm{exc,0} = 1.
\end{align}
A2: $A_\textrm{exc,0}$ is no longer a global variable extracted by fitting all curves, but within each decade of $V_{\textrm{AC}}$., see Eq. \eqref{Aexc_decade}, same as the assumption (2).\\

\noindent A3: $A_\textrm{exc,0}$ is no longer a global variable extracted by fitting all curves, but within each decade of $p_i$, here $k \in (\lfloor \textrm{lg}(p_i) \rfloor - \lfloor \textrm{lg}(p_{i,\textrm{min}}) \rfloor)$:
\begin{align}
 A_{\textrm{exc},k} = \sum_{i=1}^{N_k} W_{i,k}\cdot \frac{|A|_{i,k}(\omega)}{A_{\textrm{memb},k} V_{\textrm{AC}i,k}},~~ 
 A_{\textrm{exc,0},k} = \frac{A_{\textrm{exc},k} }{\langle A_{\textrm{exc},k}(\omega',...) \rangle_{\omega'}} \;.%\label{Aexc_decade}
\end{align}
A4: $A_\textrm{exc,0}$ is intrinsically amplitude dependent. The correction is approximated by a power law of second order $1+bA_\textrm{memb}+aA_\textrm{memb}^2$.
\begin{align}
 &A_\textrm{exc} = \sum_{i=1}^N W_i\cdot \frac{|A|_i(\omega)}{A_\textrm{memb} V_{\textrm{AC}i} \cdot (1+bA_\textrm{memb}+aA_\textrm{memb}^2)},~~ 
 A_{\textrm{exc,0}} = \frac{A_{\textrm{exc}} }{\langle A_{\textrm{exc}}(\omega',...) \rangle_{\omega'}} \;.%\label{Aexc_n1}
\end{align}
A5: $A_\textrm{exc,0}$ is drive amplitude dependent. The correction is approximated by a power law of second order. See Eq \ref{Aexc_n1}, same as the assumption (1). \\

\indent Except for the assumption A4, all other models show their potential correctness and can be further tested in the combined constrained fitting procedure. \\
\indent In Fig. \ref{fig:SM_constraints_EM08} (c),  we individually tested the following assumptions of the physical model for the coupling strength $S_i$. \\

\noindent S1: $S_i$ depends linearly on $V_\textrm{AC}$:
\begin{align}
S_i = S_0 (1 + aV_\textrm{AC})
 \end{align}
S2: $S_i$ are fixed in each decade of $V_\textrm{AC}$ but are allowed to vary from decade to decade, same as assumption (3), see Eq. \eqref{S1_decade}.\\

\noindent S3: $S_i$ are excitation dependent following a power law of $2^{nd}$ order:
\begin{align}
 S_{i} = S_{i,0}\cdot(1 + S_{i,1}\cdot V_\textrm{AC} + S_{i,2}\cdot V_\textrm{AC}^2). \label{S1_poly2_Volt}
\end{align}
S4: $S_i$ are excitation dependent following a power law with $3^{rd}$ order, see Eq. \eqref{S1_poly3_Volt}, same as assumption (4).\\

\noindent S5: $S_i$ are excitation dependent following a exponential relation.
\begin{align}
 S_{i} = S_{i,0} \cdot (1 + aV_\textrm{AC}^b). \label{S1_expo_Volt}
\end{align}
S6: $S_i$ are excitation and pressure dependent following the relation of:
\begin{align}
 S_{i} = S_{i,0} \cdot (1 + ap_i) \cdot V_{\textrm{AC},i}^b. \label{S1_lin_mbar_expo_Volt}
\end{align}
\indent Here the assumption S2 shows small deviations and hence seems to be appropriate. Thus, we consider S2 as the constraint for the assumption of $S_i$ in the combined constrained fitting procedure of C3. The assumption S4 is also used in procedure C4, which accounts for the failure of the C4 model.\\%
\indent In Fig. \ref{fig:SM_constraints_EM08} (d), we individually tested the assumption of the physical model for $\beta$ and $f_0$.\\

\noindent F1: $f_0$ depends linearly on the pressure for $p > 1.5$ mbar, same as assumption (5). \\

\noindent B1: $\beta$ depends linearly on the pressure, same as assumption (6). \\ 

\indent The assumption F1 shows almost no deviation from the free-fitting procedure corroborating its correctness. The constraint B1 results in a systematic deviation compared to the free-fitting results. This indicates that the assumption of a linear dependence of $\beta$ on $p$ $<$ 1.0 mbar is not appropriate in the present case. The deviation is strongest in the low-pressure range suggesting that corrections for the low-pressure range should be considered, which is also one of the reasons that C1 and C2 deviate from the free-fitting results.\\

%%%%%%%%%%%%%%%%%%%%%%%%%%%%%%%%%%%%%%%%%%%%%%%%%%
\section{The universality of the free and constrained fitting method tested by VICL measurements}
%------------------------------------
\subsection{Different modes and temperature effect in the fitting procedure}
\begin{figure*}
  \centering
  \includegraphics[width=\linewidth]{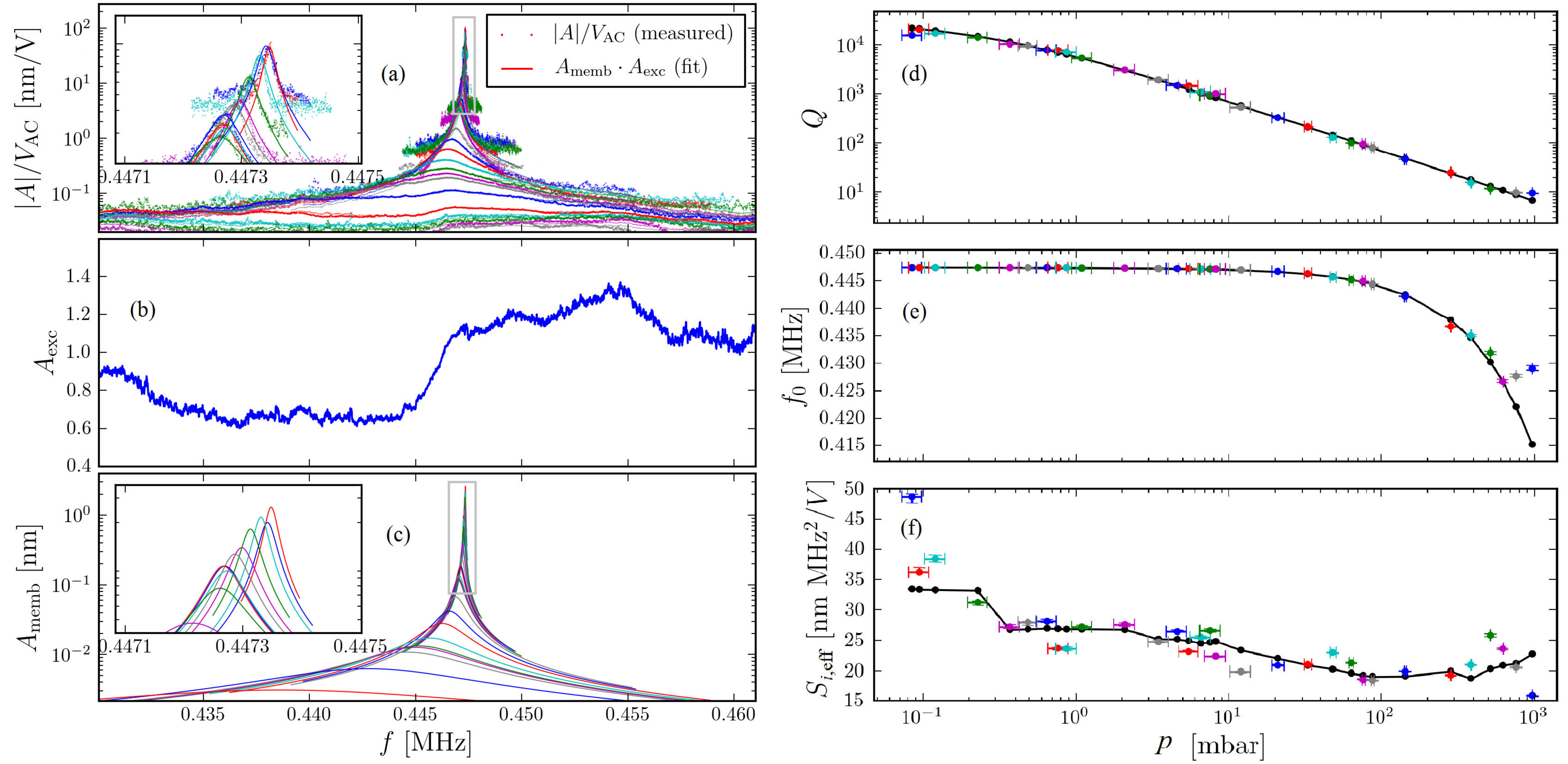}
  \caption{Fitted VICL result of the ground mode (1,1) vibration of sample OM2 at 0.447 MHz. The measured resonance curves under different pressures with their individual fitting are plotted in (a). The separated $A_\textrm{exc}$ and $A_\textrm{memb}$ are plotted in (b) and (c), respectively. (d) Fitted curve of the $Q$ factor of the ground mode (1,1) vibration of the $SiN$ membrane at 0.447 MHz. (e) Eigenfrequency of the corresponding undamped  oscillator $f_0$. (f) Proportionality factors $S_{i,\textrm{eff}}$ corresponding to the coupling strength between the excitation oscillation and the membrane oscillation. Black dots and lines plotted in subfigure (d - f) represent the corresponding constrained fitting results.}
  \label{fig:vicl_490kHz_fit}
\end{figure*}

\begin{figure*}
  \centering
  \includegraphics[width=\linewidth]{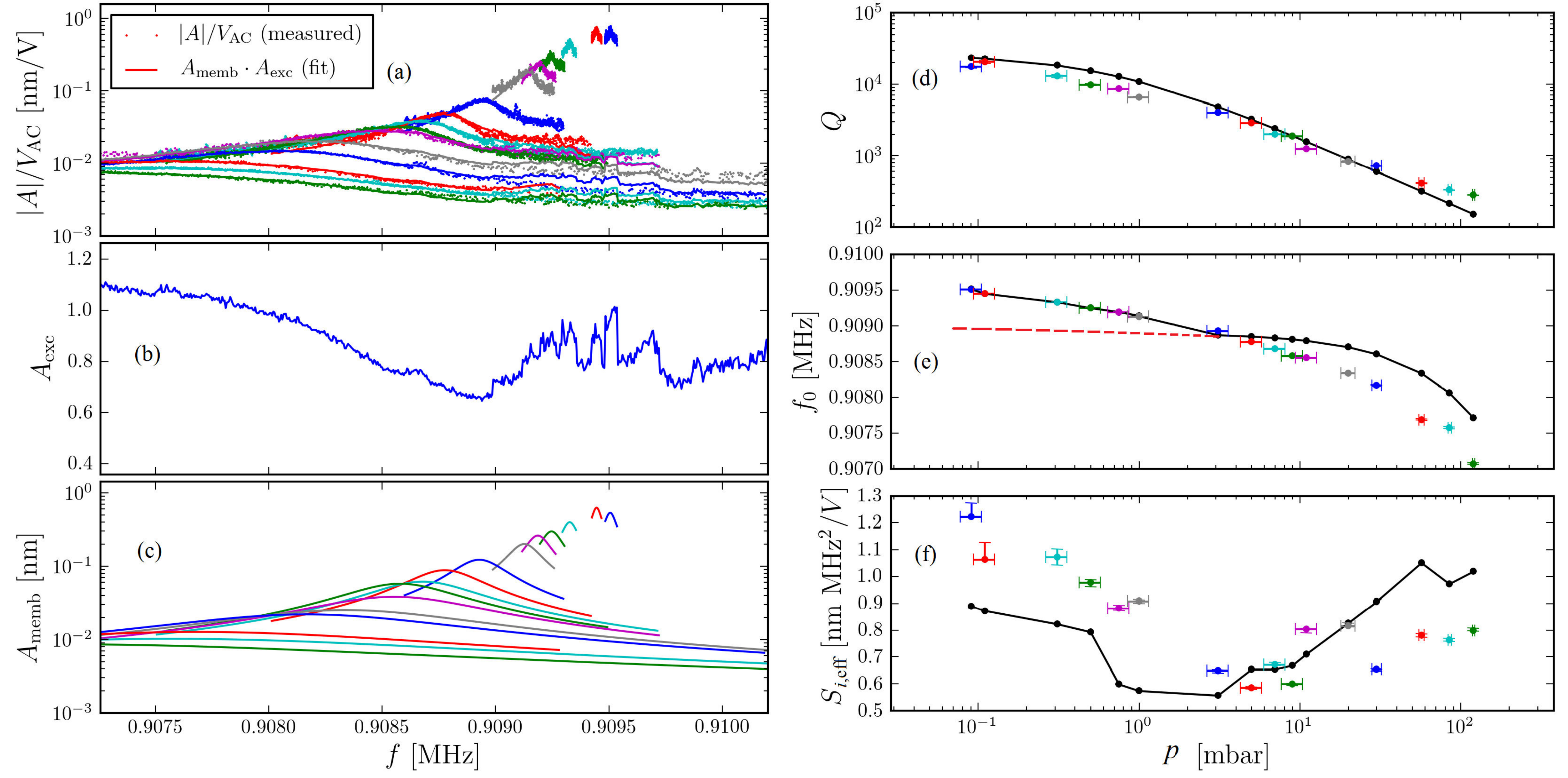}
  \caption{VICL results: fitted of the (2,2) mode vibration of sample OM2 at 0.909 MHz. The measured resonance curves under different pressures with their individual fitting are plotted in (a). The separated $A_\textrm{exc}$ and $A_\textrm{memb}$ are plotted in (b) and (c), respectively. (d) Fitted curve of the $Q$ factor of the (2,2) mode vibration of sample OM2 at 0.909 MHz. (e) Eigenfrequency of the corresponding undamped  oscillator $f_0$. (f) Proportionality factors $S_{i,\textrm{eff}}$ corresponding to the coupling strength between excitation oscillation and membrane oscillation. Black dots and lines plotted in subfigure (d - f) represent the corresponding constrained fitting results.}
  \label{fig:vicl_900kHz_fit}
\end{figure*}

\noindent To underline the general applicability of the free and constrained fitting method, we now employ it to the VICL optical measured results of the (1,1) and (2,2) mode of a forth sample, OM2, a 478 nm thick SiN membrane with a lateral size of 298$\times$296 $\mu$m$^2$.\\
\indent Figure \ref{fig:vicl_490kHz_fit} shows the results for the ground mode (labeled the (1,1) mode, because it has one maximum in $x$ and one in $y$ direction) of sample OM2. In Fig. \ref{fig:vicl_490kHz_fit}(a) and (c), each curve represents a specific pressure: a lower pressure gives rise to a sharper peak with a higher amplitude. We also observe a slight shift of the resonance to higher frequencies with decreasing pressure in the low-pressure range, in contrast to the behavior of a damped harmonic oscillator. At very low pressure even non-monotonic behavior of the resonance frequency on the pressure may occur. We will comment on these two effects in the following description. The frequency dependence of the amplitude of the excitation system $A_\textrm{exc}$ is presented in panel (b) of Fig. \ref{fig:vicl_490kHz_fit}. The fit results  of the $Q$ factor in Fig. \ref{fig:vicl_490kHz_fit}(d) increase with decreasing pressure as expected, if the damping is dominated by the air environment. At low pressure, the $Q$ factor saturates, indicating that the intrinsic damping of the membrane starts to dominate. From the extrapolation to low pressure the intrinsic $Q$ factor and the eigenfrequency of the ground mode (1,1) are estimated as $Q=2\times 10^4$ and $f_0=0.447$ MHz, respectively.  By comparing with the theoretical eigenfrequencies expression: $f_{m,n} \simeq \sqrt{\sigma(m^2+n^2)/(4\rho L^2)}$, where $\sigma$, $\rho$, $L$ and $(m,n)$ correspond to the tensile stress, mass density, lateral size of the membrane and the integer mode indices representing the number of antinodes, respectively \cite{chakram2014dissipation, yu2012control}, with the values $\sigma = 1.099\times 10^{-1}$ GPa (measured by our phase shifting and surface amplitude method \cite{waitz2012mode}. Here the $\rho = 3180$ kg/cm$^3$, and $L \simeq 3.0\times 10^{-4}$ m, for the (1,1) mode, we obtain $f_{1,1} \simeq 0.438$ MHz, which is $2.0\%$ lower than our fitting results.\\
\indent Now we discuss the excited (2,2) mode which has resonance frequencies around 900 kHz. Curves measured in a pressure range from 120 mbar to 0.091 mbar are plotted in Fig. \ref{fig:vicl_900kHz_fit}(a). Due to the relatively small amplitude of this mode, the limited SNR, and the very large width of the resonance curves, measurements at higher pressures are not possible. Similarly, at low pressure, very small excitation amplitudes have to be used to stay in the linear oscillator regime. When increasing the excitation, the shape of the resonance curves starts to deviate from the Lorentz curve shape. As a result the accessible frequency range is limited. As before, the contributions of the membrane and the excitation system are separated by the VICL analysis. The eigenfrequency of the (2,2) mode is estimated as $f_0=0.909$ MHz. For the (2,2) mode, $f_{2,2} \simeq 0.876$ MHz can be determined by the theoretical eigenfrequency equation with a discrepancy of $3.72\%$ compared with the fitting results.\\
\indent Figure \ref{fig:vicl_900kHz_fit}(e) shows a significant shift of $f_0$ of the (2,2) mode in higher frequencies when the pressure is lowered. This blue shift amounts to roughly 0.5 kHz below $p = 3$ mbar and is much more pronounced than for the (1,1) mode that increases by roughly 0.1 kHz in the same frequency range. In Fig. \ref{fig:vicl_900kHz_fit}(d), the low-pressure limit of the $Q$ factor of the membrane is estimated to be $Q=2\times 10^4$, in agreement with the results for the (1,1) mode and other modes for the same membrane (not shown). It also fits well with the observed scaling with $(L/h)^2$ reported in Ref.~\citenum{chakram2014dissipation}, from where we extrapolate our membrane to have a $Q$ factor around $1.5\times 10^4$. This agreement indicates that the leveling off of $Q$ at low pressure signals indeed the onset of intrinsic damping. The decrease of $Q$ with increasing pressure for the (2,2) mode is weaker than for the (1,1) mode, such that in general the (2,2) mode has a higher $Q$ factor than the (1,1) mode, indicating that the viscous damping of the environment is more effective for the ground mode. This behaviour can be understood since the integral displacement in the ground mode is much bigger than in the excited (2,2) mode. \\
\indent We now turn to the constrained fitting. The fitting results of the constrained model (black dots and lines in the right panels of Fig. \ref{fig:vicl_490kHz_fit} and \ref{fig:vicl_900kHz_fit}) approximately follow those of the free-fitting model. However, the fitting results for the eigenfrequency of the (2,2) mode in the higher pressure range, as shown in Fig. \ref{fig:vicl_900kHz_fit}(e),  deviate systematically from the free-fitting results, indicating that constraint (2) does not accurately describe the experiment in that range, probably due to the mechanism described above. Another contributing possibility would be an enhancement of the effective mass of the membrane due to the inertia of the surrounding atmosphere, as observed earlier for Si membranes \cite{waitz2012mode}. The $S_{i,\textrm{eff}}$ presented in panel (f) of Fig. \ref{fig:vicl_490kHz_fit} and \ref{fig:vicl_900kHz_fit} show a relatively weak dependence on pressure, supporting the harmonic oscillator model, in which a constant $S$ is expected.  For the (1,1) mode the constrained fit results agree well with the free-fitting results, except for the two data points at the extremes of the pressure range. For the (2,2) mode the $S_{i,\textrm{eff}}$ values of the free and the constrained model both scatter around a constant value, and the  deviation between freely-fitted results and constrained results is smaller than the scatter. We attribute the scatter to the small overall amplitude which is  a factor of 25 smaller than for the (1,1) mode.\\
\indent We now turn to a possible explanation of the frequency shifts and increased scattering, especially in the lower pressure range, which are directly presented in Fig. \ref{fig:vicl_900kHz_fit}(c) and \ref{fig:vicl_900kHz_fit}(e). The $f_0$ in the weak damping regime at low pressure should be approximately constant according to the expression of the damped driven harmonic oscillator in Eq. (\ref{Amemb}). 
Possible reasons for the frequency shift include the following: When running a complete VICL test at room temperature which takes about 3 hours under continuous heat input, at low pressure the temperature might not be constant. Further potential error sources are the finite precision of the pressure measurement and the stability of the electromechanical performance of the piezo ring. However, the pressure sensors are well calibrated and the contribution of the excitation system including the piezo ring is separated and extracted by our analysis. Thus we argue that the temperature variation might cause a slight shift of the eigenfrequency.\\
\indent In Fig. \ref{fig:vicl_900kHz_fit}(a) and \ref{fig:vicl_900kHz_fit}(c), the frequency shift can be directly observed. The curves have been recorded for decreasing temperature. From 2 mbar down to 0.01 mbar $f_0$ increases by roughly 500 Hz. The temperature sensitivity indicated above was determined for the (1,1) mode. When assuming that the temperature change of $f_0$ scales with $f_0$, we imply the temperature dependent resonance frequency measurement on (2,2) mode and we find the temperature sensitivity of the (2,2) mode of approximately 113 Hz/$^{\circ}\mathrm{C}$, corresponding to a temperature increase of $4.4\,^{\circ}\mathrm{C}$.  Such a temperature increase can well be caused by the heat dissipation of the running equipment of the VICL test. We therefore attribute the deviations of these low-pressure data points to slight increases of the temperature. A constant temperature of the sample would result in the red dashed line in Fig. \ref{fig:vicl_900kHz_fit}(e). \\
%
%-------------------------------------------------------
\subsection{Temperature-dependent resonance frequencies} 
\begin{figure}
  \centering
  \includegraphics[width=0.6\linewidth]{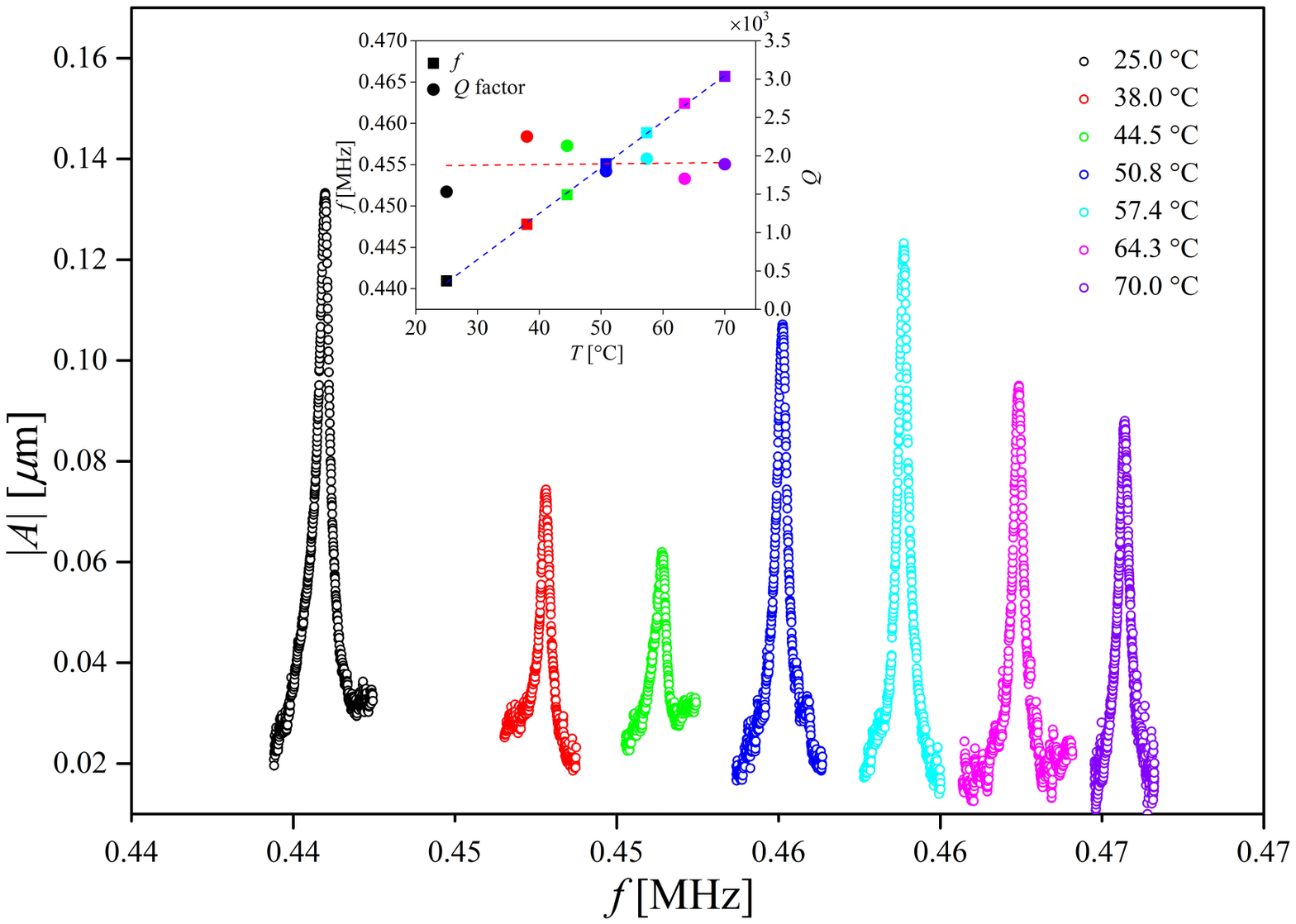}
  \caption{Resonance curves of the ground mode vibration at different temperatures of sample OM2. Curves are measured in a narrow pressure range of 3-4 mbar. Temperatures are directly measured by the sensor which is fixed in the copper stage and next to the sample. Curves are measured at $25.0\,^{\circ}\mathrm{C}$, $38.0\,^{\circ}\mathrm{C}$, $44.5\,^{\circ}\mathrm{C}$, $50.8\,^{\circ}\mathrm{C}$, $57.4\,^{\circ}\mathrm{C}$, $64.3\,^{\circ}\mathrm{C}$ and $70.0\,^{\circ}\mathrm{C}$. The values of the resonance frequencies and $Q$ factors at different temperatures are plotted in the inserted image. The colors of the dots correspond to those of the resonance curves. The values of the resonance frequencies and $Q$ factors follow linear tendencies illustrated by the linear fitting result (blue and red dashed line). The error bars are smaller than the symbol size.}
  \label{fig:VICL_of_temperature_dependence_eigen_frequency}
\end{figure}

\noindent The temperature variation might cause a slight shift of the eigenfrequency, to quantify this effect we study the sensitivity of the response curves of the ground mode of sample OM2 to temperature changes, as shown in Fig. \ref{fig:VICL_of_temperature_dependence_eigen_frequency}. We placed the sample holder on  an auto-controlled heating stage to adjust the temperature of the sample. The resolution of the temperature control stage is $1\,^{\circ}\mathrm{C}$. The heating stage is placed under the vacuum chamber and heats the whole vacuum stage to the set temperature. Temperatures are directly measured by the sensor which is fixed in the copper stage and next to the sample. Since the interference pattern is highly sensitive to any changes of the geometry caused by thermal nonequilibrium (thermal expansion, temperature gradients, etc.), we use the camera of the interferometer to monitor the thermalization process, which may take up to 2 hours. When the interference pattern is stable in time we start the VICL measurements.\\
\indent The curves have been measured for ascending temperatures in a moderate vacuum range of 3-4 mbar where the SNR is highest. The minimum setting temperature of the heating stage is $38.0\,^{\circ}\mathrm{C}$. The resonance frequency shift between the lowest and the highest temperature is about 3.5 kHz with an approximately linear increasing relation between the temperature and the resonance frequency, presented in the inserted image of Fig.  \ref{fig:VICL_of_temperature_dependence_eigen_frequency}. Thus the resonance frequency shift can be averaged as 550 Hz/$^{\circ}\mathrm{C}$. This increase may arise from the different thermal expansion coefficients of the silicon chip and the SiN membrane \cite{sinha1978thermal} and will be studied in more detail in a forthcoming publication.\\
\indent Figure \ref{fig:VICL_of_temperature_dependence_eigen_frequency} also shows that the $Q$ factor is almost independent of temperature under these conditions. We deduce an average value of $1.9\pm 0.4 \times 10^3$ agreeing well with the value observed at room temperature, as shown in Fig. \ref{fig:vicl_490kHz_fit}(d). We conclude that in this temperature range the viscosity of the air and its resulting damping are constant. When operating the membrane resonator under these conditions, the FWHM of the ground mode amounts to approximately 200 Hz, meaning that temperature changes of 0.4 K are detectable. This high temperature sensitivity makes SiN membrane resonators suitable as temperature sensors. For lower pressures and even higher temperature resolution is possible because of the higher $Q$-factor. For example, in Fig. \ref{fig:vicl_900kHz_fit}(c),  the resonance frequencies of the lowest two pairs of pressure values are shifted by 50 Hz and 130 Hz, respectively, corresponding to a temperature variation of $0.44\,^{\circ}\mathrm{C}$ and $1.15\,^{\circ}\mathrm{C}$, respectively.\\

\bibliographystyle{apsrev4-1}
\bibliography{refs_air}